\documentclass[aps,prd,twocolumn,superscriptaddress,nofootinbib]{revtex4-1}

\usepackage{latexsym}
\usepackage{amsmath}
\usepackage{amssymb}
\usepackage{amsfonts}
\usepackage{bm}

\usepackage{color}

\usepackage{supertabular} 
\usepackage{placeins}
\usepackage{epsfig}
\usepackage{graphicx}

\begin{document}


\title{Hidden-charm tetraquarks with strangeness in the chiral quark model}


\author{Gang Yang}
\email[]{yanggang@zjnu.edu.cn}
\affiliation{Department of Physics, Zhejiang Normal University, Jinhua 321004, China}

\author{Jialun Ping}
\email[]{jlping@njnu.edu.cn}
\affiliation{Department of Physics and Jiangsu Key Laboratory for Numerical Simulation of Large Scale Complex Systems, Nanjing Normal University, Nanjing 210023, P. R. China}

\author{Jorge Segovia}
\email[]{jsegovia@upo.es}
\affiliation{Departamento de Sistemas F\'isicos, Qu\'imicos y Naturales, \\ Universidad Pablo de Olavide, E-41013 Sevilla, Spain}



\begin{abstract}
The hidden-charm tetraquarks with strangeness, $c\bar{c}s\bar{q}$ $(q=u,\,d)$, in $J^P=0^+$, $1^+$ and $2^+$ are systematically investigated in the framework of real- and complex-scaling range of a chiral quark model, whose parameters have been fixed in advance describing hadron, hadron-hadron and multiquark phenomenology. Each tetraquark configuration, compatible with the quantum numbers studied, is taken into account; this includes meson-meson, diquark-antidiquark and K-type arrangements of quarks with all possible color wave functions in four-body sector. Among the different numerical techniques to solve the Schr\"odinger-like 4-body bound state equation, we use a variational method in which the trial wave function is expanded in complex-range Gaussian basis functions, which is characterized by its simplicity and flexibility. This theoretical framework has already been used to study different kinds of multiquark systems, such as the hidden-charm pentaquarks, $P^+_c$, and doubly-charmed tetraquarks, $T^+_{cc}$. The recently reported $Z_{cs}$ states by the BESIII and LHCb collaborations are generally compatible with either compact tetraquark or hadronic molecular resonance configurations in our investigation. Moreover, several additional exotic resonances are found in the mass range between 3.8 GeV and 4.6 GeV.
\end{abstract}

\pacs{
12.38.-t \and 
12.39.-x      
}
\keywords{
Quantum Chromodynamics \and
Quark models
}

\maketitle


\section{Introduction}

A structure with a significance of $5.3\sigma$ was reported in the process of $e^+e^-\rightarrow K^+(D^-_s D^{*0}+D^{*-}_s D^0)$ by the BESIII collaboration~\cite{BESIII:2020qkh}, its experimentally measured mass and width were $3982.5^{+1.8}_{-2.6}\pm 2.1$ MeV and $12.8^{+5.3}_{-4.4}\pm 3.0$ MeV, respectively. Undoubtedly, the named $Z_{cs}(3985)$ state was the first candidate of a charged hidden-charm tetraquark state with strangeness. 

Later on, more charmonium-like states with strange content were reported by the LHCb collaboration in proton-proton collisions~\cite{LHCb:2021uow}. The $Z_{cs}(4000)$ is observed in the $B^+\to J/\psi \phi K^+$ decay, with mass and width $4003\pm 6^{+4}_{-14}$ MeV and $131\pm 15\pm 26$ MeV, and the preferred spin-parity is $J^P=1^+$. The $X(4685)$, also with $J^P=1^+$ quantum numbers, decays to $J/\psi \phi$ final state with a high significance claimed by the collaboration. Furthermore, the $Z_{cs}(4220)$ and $X(4630)$ are also reported with significance exceeding 5$\sigma$ derivations. These facts trigger many theoretical investigations on the nature of hidden-charm tetraquark with strangeness.

In many theoretical works the $Z_{cs}(3985)$ is identified as the strange partner of the $Z_c(3900)$ within the $SU(3)_f$ symmetry, and thus the hadronic molecular configuration is proposed. In particular, the $D^*\bar{D}_s-D\bar{D}^*_s$ and $D^*\bar{D}^*_s$ molecules with spin-parity $J^P=1^+$ state can be related to the mentioned $Z^{(\ast)}_{cs}$ particles~\cite{Yang:2020nrt}. This result is also supported by a coupled-channel calculation~\cite{Meng:2020ihj}, a variety of effective field theory frameworks~\cite{Sun:2020hjw, Wang:2020htx, Ikeno:2020mra, Ding:2021igr}, approaches based on QCD sum rules~\cite{Wang:2020rcx, Xu:2020evn} and potential model descriptions~\cite{Yan:2021tcp}. Finally, within the framework of an effective range expansion, a unified description of the hidden-charm tetraquark states $Z_c(3900)$, $Z_{cs}(3985)$ and $X(4020)$ are discussed~\cite{Guo:2020vmu}. Meanwhile, the 2- and 4-body configuration mixing scheme for describing the $Z_{cs}(3985)$ state has been proposed to be crucial in many theoretical investigations, \emph{viz.} the $Z_{cs}(3985)$ is excluded as a pure $D^{*0}D^-_s/D^0D^{*-}_s/D^{*0}D^{*-}_s$ hadronic molecular state in, for instance, one-boson-exchange model~\cite{Chen:2020yvq, Liu:2020nge} and constituent quark model~\cite{Jin:2020yjn}. The $Z_{cs}(3985)$ and $Z_{cs}(4003)$ can be explained well within a mixture formalism in Refs.~\cite{Karliner:2021qok, Wan:2020oxt}. 

Notwithstanding this, many theoretical approaches conclude that a compact tetraquark structure is also possible for the $Z_{cs}(3985)$, \emph{e.g.} one-boson-exchange model~\cite{Liu:2020nge}, constituent quark model~\cite{Jin:2020yjn} and QCD sum rules~\cite{Wang:2020iqt}. Furthermore, some novel pictures for the $Z_{cs}(3985)$ state are proposed. Particularly, the $Z_{cs}(3985)$ can be identified as a reflection structure of charmed-strange meson $D^*_{s2}(2573)$~\cite{Wang:2020kej}. It is also explained as a genuine state, either virtual or bound, in a contact potential model~\cite{Du:2020vwb}. Additionally, the photo-production~\cite{Cao:2020cfx} and properties of $Z_{cs}(3985)$ in hot dense medium~\cite{Azizi:2020zyq, Sungu:2020zvk} have been recently studied theoretically.

Concerning the $Z_{cs}(4000)$, $Z_{cs}(4220)$, $X(4630)$ and $X(4685)$ states, there are in the literature interpretations compatible with hadronic molecules~\cite{Chen:2021erj, Meng:2021rdg, Yang:2021sue, Wang:2021ghk}, compact tetraquark structures~\cite{Chen:2021uou, Giron:2021sla, Wang:2021ghk, Turkan:2021ome} and even non-resonance configurations~\cite{Ge:2021sdq}. Besides, the magnetic moments of the $Z_{cs}(4000)$ and $Z_{cs}(4220)$ are calculated by means of light-cone QCD sum rules~\cite{Ozdem:2021hka}.

In order to disentangle the nature of these charmonium-like resonances with strangeness announced recently by the LHCb and BESIII collaborations, a systematical investigation on the hidden-charm tetraquarks with strange content: $c\bar{c}s\bar{q}$ $(q=u,\,d)$, is performed within a chiral quark model formalism. The same theoretical framework has already been applied with success in the description of other multiquark systems, \emph{e.g.}, hidden-charm and -bottom pentaquarks~\cite{Yang:2015bmv, Yang:2018oqd}, doubly-charm pentaquarks~\cite{gy:2020dcp}, doubly-heavy tetraquarks, $QQ\bar{q}\bar{q}$~\cite{gy:2020dht, gy:2020dhts} and strange-heavy tetraquarks, $sQ\bar{q}\bar{q}$ $(q=u,\,d,\,s;\,Q=c,\,b)$~\cite{Yang:2021izl}. Particularly, we have explained the hidden-charm pentaquarks~\cite{Yang:2015bmv}, $P^+_c(4312)$, $P^+_c(4380)$, $P^+_c(4440)$ and $P^+_c(4457)$, reported by the LHCb collaboration~\cite{Aaij:2015tga, lhcb:2019pc}, and predicted the doubly charmed tetraquark~\cite{gy:2020dht}, $T^+_{cc}$, announced very recently by the same experimental collaboration~\cite{LHCb:2021vvq, LHCb:2021auc}. It is also worth highlighting that the same theoretical approach was previously applied to the charmonium, bottomonium and heavy baryon sectors, studying their spectra~\cite{Segovia:2008zz, Segovia:2013wma, Segovia:2016xqb, Yang:2019lsg}, their electromagnetic, weak and strong decays and reactions~\cite{Segovia:2009zz, Segovia:2011zza, Segovia:2012cd, Segovia:2015dia}, and their coupling with meson-meson thresholds~\cite{Ortega:2009hj, Ortega:2016mms, Ortega:2016pgg, Ortega:2020uvc}.

Our formulation in real- and complex-scaling method of the theoretical formalism has been discussed in detail in Ref.~\cite{YangSym2020}. The complex-scaling method (CSM) allows us to distinguish three kinds of scattering singularities: bound, resonance and scattering; which allows us to perform a complete analysis of the scattering problem within the same formalism. Furthermore, the meson-meson, diquark-antidiquark and K-type configurations, plus their couplings, are considered for the tetraquark system. Finally, the Rayleigh-Ritz variational method is employed in dealing with the spatial wave functions of the $c\bar{c}s\bar{q}$ tetraquark states, which are expanded by means of the well-known Gaussian expansion method (GEM) of Ref.~\cite{Hiyama:2003cu}.

The manuscript is arranged as follows. In Sec.~\ref{sec:model} the theoretical framework is presented; we briefly describe the complex-range method applied to a chiral quark model and the $c\bar{c}s\bar{q}$ $(q=u,\,d)$ tetraquark wave-functions. Section~\ref{sec:results} is devoted to the analysis and discussion of the obtained low-lying $c\bar{c}s\bar{q}$ $(q=u,\,d)$ tetraquark states with $J^P=0^+$, $1^+$ and $2^+$, and isospin $I=1/2$. Finally, we summarize and give some prospects in Sec.~\ref{sec:summary}.


\section{Theoretical framework}
\label{sec:model}

A throughout review of the theoretical formalism used herein has been recently published in Ref.~\cite{YangSym2020}. We shall, however, focused on the most relevant features of the chiral quark model and the numerical method concerning the strange hidden-charm tetraquarks, \emph{viz.} the $c\bar{c}s\bar{q}$ system.
 
Within the so-called complex-range investigations, the relative coordinate of a two-body interaction is rotated in the complex plane by an angle $\theta$, \emph{i.e.} $\vec{r}_{ij}\to \vec{r}_{ij} e^{i\theta}$. Therefore, the general form of the four-body Hamiltonian reads:
\begin{equation}
H(\theta) = \sum_{i=1}^{4}\left( m_i+\frac{\vec{p\,}^2_i}{2m_i}\right) - T_{\text{CM}} + \sum_{j>i=1}^{4} V(\vec{r}_{ij} e^{i\theta}) \,,
\label{eq:Hamiltonian}
\end{equation}
where $m_{i}$ is the quark mass, $\vec{p}_i$ is the quark's momentum, and $T_{\text{CM}}$ is the center-of-mass kinetic energy. According to the so-called ABC theorem~\cite{JA22269, EB22280}, the complex scaled Schr\"odinger equation:
\begin{equation}\label{CSMSE}
\left[ H(\theta)-E(\theta) \right] \Psi_{JM}(\theta)=0	
\end{equation}
has (complex) eigenvalues which can be classified into three different kinds: bound, resonance and continuum (scattering) states. Those which are either bound or resonance are independent of the rotated angle $\theta$; however, the first ones are always fixed on the coordinate-axis (there is no imaginary part of the eigenvalue), whereas the second ones are located above the corresponding threshold lines with a total decay width $\Gamma=-2\,\text{Im}(E)$.

The dynamics of the $c\bar{c}s\bar{q}$ tetraquark system is driven by a two-body potential
\begin{equation}
\label{CQMV}
V(\vec{r}_{ij}) = V_{\chi}(\vec{r}_{ij}) + V_{\text{CON}}(\vec{r}_{ij}) + V_{\text{OGE}}(\vec{r}_{ij})  \,,
\end{equation}
which takes into account the most relevant features of QCD at its low energy regime: dynamical chiral symmetry breaking, confinement and the perturbative one-gluon exchange interaction. Herein, the low-lying $S$-wave positive parity $c\bar{c}s\bar{q}$ tetraquark states shall be investigated, and thus the central and spin-spin terms of the potential are the only ones needed.

One consequence of the dynamical breaking of chiral symmetry is that Goldstone boson exchange interactions appear between constituent light quarks $u$, $d$ and $s$. Therefore, the chiral interaction can be written as~\cite{Vijande:2004he}:
\begin{equation}
V_{\chi}(\vec{r}_{ij}) = V_{\pi}(\vec{r}_{ij})+ V_{\sigma}(\vec{r}_{ij}) + V_{K}(\vec{r}_{ij}) + V_{\eta}(\vec{r}_{ij}) \,,
\end{equation}
given by
\begin{align}
&
V_{\pi}\left( \vec{r}_{ij} \right) = \frac{g_{ch}^{2}}{4\pi}
\frac{m_{\pi}^2}{12m_{i}m_{j}} \frac{\Lambda_{\pi}^{2}}{\Lambda_{\pi}^{2}-m_{\pi}
^{2}}m_{\pi} \Bigg[ Y(m_{\pi}r_{ij}) \nonumber \\
&
\hspace*{1.20cm} - \frac{\Lambda_{\pi}^{3}}{m_{\pi}^{3}}
Y(\Lambda_{\pi}r_{ij}) \bigg] (\vec{\sigma}_{i}\cdot\vec{\sigma}_{j})\sum_{a=1}^{3}(\lambda_{i}^{a}
\cdot\lambda_{j}^{a}) \,, \\
& 
V_{\sigma}\left( \vec{r}_{ij} \right) = - \frac{g_{ch}^{2}}{4\pi}
\frac{\Lambda_{\sigma}^{2}}{\Lambda_{\sigma}^{2}-m_{\sigma}^{2}}m_{\sigma} \Bigg[Y(m_{\sigma}r_{ij}) \nonumber \\
&
\hspace*{1.20cm} - \frac{\Lambda_{\sigma}}{m_{\sigma}}Y(\Lambda_{\sigma}r_{ij})
\Bigg] \,,
\end{align}
\begin{align}
& 
V_{K}\left( \vec{r}_{ij} \right)= \frac{g_{ch}^{2}}{4\pi}
\frac{m_{K}^2}{12m_{i}m_{j}}\frac{\Lambda_{K}^{2}}{\Lambda_{K}^{2}-m_{K}^{2}}m_{
K} \Bigg[ Y(m_{K}r_{ij}) \nonumber \\
&
\hspace*{1.20cm} -\frac{\Lambda_{K}^{3}}{m_{K}^{3}}Y(\Lambda_{K}r_{ij}) \Bigg] (\vec{\sigma}_{i}\cdot\vec{\sigma}_{j}) \sum_{a=4}^{7}(\lambda_{i}^{a} \cdot \lambda_{j}^{a}) \,, \\
& 
V_{\eta}\left( \vec{r}_{ij} \right) = \frac{g_{ch}^{2}}{4\pi}
\frac{m_{\eta}^2}{12m_{i}m_{j}} \frac{\Lambda_{\eta}^{2}}{\Lambda_{\eta}^{2}-m_{
\eta}^{2}}m_{\eta} \Bigg[ Y(m_{\eta}r_{ij}) \nonumber \\
&
\hspace*{1.20cm} -\frac{\Lambda_{\eta}^{3}}{m_{\eta}^{3}
}Y(\Lambda_{\eta}r_{ij}) \Bigg] (\vec{\sigma}_{i}\cdot\vec{\sigma}_{j})
\Big[\cos\theta_{p} \left(\lambda_{i}^{8}\cdot\lambda_{j}^{8}
\right) \nonumber \\
&
\hspace*{1.20cm} -\sin\theta_{p} \Big] \,,
\end{align}
where $Y(x)=e^{-x}/x$ is the standard Yukawa function. The physical $\eta$ meson, instead of the octet one, is considered by introducing the angle $\theta_p$. The $\lambda^{a}$ are the SU(3) flavor Gell-Mann matrices. Taken from their experimental values, $m_{\pi}$, $m_{K}$ and $m_{\eta}$ are the masses of the SU(3) Goldstone bosons. The value of $m_{\sigma}$ is determined through the PCAC relation $m_{\sigma}^{2}\simeq m_{\pi}^{2}+4m_{u,d}^{2}$~\cite{Scadron:1982eg}. Finally, the chiral coupling constant, $g_{ch}$, is determined from the $\pi NN$ coupling constant through
\begin{equation}
\frac{g_{ch}^{2}}{4\pi}=\frac{9}{25}\frac{g_{\pi NN}^{2}}{4\pi} \frac{m_{u,d}^{2}}{m_{N}^2} \,,
\end{equation}
which assumes that flavor SU(3) is an exact symmetry only broken by the different mass of the strange quark. Herein, we should notice that only one $\bar{q}$ $(q=u,\,d)$ and one $s$ light quark is considered in the tetraquark system, hence the $\pi$-meson exchange potential will be excluded in the chiral interaction.

Color confinement should be encoded in the non-Abelian character of QCD. It has been demonstrated by lattice-regularized QCD that multi-gluon exchanges produce an attractive linearly rising potential proportional to the distance between infinite-heavy quarks~\cite{Bali:2005fu}. However, the spontaneous creation of light-quark pairs from the QCD vacuum may give rise at the same scale to a breakup of the created color flux-tube~\cite{Bali:2005fu}. These two observations can be described phenomenologically by
\begin{equation}
V_{\text{CON}}(\vec{r}_{ij})=\left[-a_{c}(1-e^{-\mu_{c}r_{ij}})+\Delta \right] 
(\lambda_{i}^{c}\cdot \lambda_{j}^{c}) \,,
\label{eq:conf}
\end{equation}
where $a_{c}$, $\mu_{c}$ and $\Delta$ are model parameters,\footnote{It is widely believed that confinement is flavor independent and thus it should be constraint by the light hadron spectra despite our aim is to determine energy states in heavier quark sectors~\cite{Segovia:2008zza, Segovia:2008zz}} and the SU(3) color Gell-Mann matrices are denoted as $\lambda^c$. One can see in Eq.~\eqref{eq:conf} that the potential is linear at short inter-quark distances with an effective confinement strength $\sigma = -a_{c} \, \mu_{c} \, (\lambda^{c}_{i}\cdot \lambda^{c}_{j})$, while it becomes constant at large distances, $V_{\text{thr.}} = (\Delta-a_c)(\lambda^{c}_{i}\cdot \lambda^{c}_{j})$.

Beyond the chiral symmetry breaking scale one expects the dynamics to be
governed by QCD perturbative effects. In particular, the one-gluon exchange potential (which includes the so-called coulomb and color-magnetic interactions) is the leading order contribution:
\begin{align}
&
V_{\text{OGE}}(\vec{r}_{ij}) = \frac{1}{4} \alpha_{s} (\lambda_{i}^{c}\cdot \lambda_{j}^{c}) \Bigg[\frac{1}{r_{ij}} \nonumber \\ 
&
\hspace*{1.60cm} - \frac{1}{6m_{i}m_{j}} (\vec{\sigma}_{i}\cdot\vec{\sigma}_{j}) 
\frac{e^{-r_{ij}/r_{0}(\mu_{ij})}}{r_{ij} r_{0}^{2}(\mu_{ij})} \Bigg] \,,
\end{align}
where $r_{0}(\mu_{ij})=\hat{r}_{0}/\mu_{ij}$ is a regulator which depends on the reduced mass of the $q\bar{q}$ pair, the Pauli matrices are denoted by $\vec{\sigma}$, and the contact term has been regularized as
\begin{equation}
\delta(\vec{r}_{ij}) \sim \frac{1}{4\pi r_{0}^{2}(\mu_{ij})}\frac{e^{-r_{ij} / r_{0}(\mu_{ij})}}{r_{ij}} \,.
\end{equation}

An effective scale-dependent strong coupling constant, $\alpha_s(\mu_{ij})$, provides a consistent description of mesons and baryons from light to heavy quark sectors. We use the definition of Ref.~\cite{Segovia:2013wma}:
\begin{equation}
\alpha_{s}(\mu_{ij})=\frac{\alpha_{0}}{\ln\left(\frac{\mu_{ij}^{2}+\mu_{0}^{2}}{\Lambda_{0}^{2}} \right)} \,,
\end{equation}
in which $\alpha_{0}$, $\mu_{0}$ and $\Lambda_{0}$ are parameters of the model.

The model parameters are listed in Table~\ref{tab:model}. Additionally, for later concern, Table~\ref{MesonMass} lists theoretical and experimental (if available) masses of $1S$ and $2S$ states of $K^{(*)}$, $D^{(*)}$, $D^{(*)}_s$, $\eta_c$ and $J/\psi$ mesons predicted within our theoretical framework.

\begin{table}[!t]
\caption{\label{tab:model} Model parameters.}
\begin{ruledtabular}
\begin{tabular}{llr}
Quark masses     & $m_q\,(q=u,\,d)$ (MeV) & 313 \\
                 & $m_s$ (MeV) &  555 \\
                 & $m_c$ (MeV) & 1752 \\[2ex]
Goldstone bosons & $\Lambda_\sigma~$ (fm$^{-1}$) &   4.20 \\
                 & $\Lambda_\eta=\Lambda_K$ (fm$^{-1}$)      &   5.20 \\
                 & $g^2_{ch}/(4\pi)$                         &   0.54 \\
                 & $\theta_P(^\circ)$                        & -15 \\[2ex]
Confinement      & $a_c$ (MeV)         & 430 \\
                 & $\mu_c$ (fm$^{-1})$ & 0.70 \\
                 & $\Delta$ (MeV)      & 181.10 \\[2ex]
OGE              & $\alpha_0$              & 2.118 \\
                 & $\Lambda_0~$(fm$^{-1}$) & 0.113 \\
                 & $\mu_0~$(MeV)           & 36.976 \\
                 & $\hat{r}_0~$(MeV~fm)    & 28.170 \\
\end{tabular}
\end{ruledtabular}
\end{table}

\begin{table}[!t]
\caption{\label{MesonMass} Theoretical and experimental (if available) masses of $nL=1S$ and $2S$ states of $K^{(*)}$, $D^{(*)}$, $D^{(*)}_s$, $\eta_c$ and $J/\psi$ mesons.}
\begin{ruledtabular}
\begin{tabular}{lccc}
Meson & $nL$ & $M_{\text{The.}}$ (MeV) & $M_{\text{Exp.}}$ (MeV) \\
\hline
$K$ & $1S$ &  $481$ & $494$ \\
    & $2S$ & $1468$ & - \\[2ex]
$K^*$ & $1S$ &  $907$ & $892$ \\
      & $2S$ & $1621$ & - \\[2ex]
$D$ & $1S$ & $1897$ & $1870$ \\
    & $2S$ & $2648$ & - \\[2ex]
$D^*$ & $1S$ & $2017$ & $2007$ \\
      & $2S$ & $2704$ & - \\[2ex]
$D_s$ & $1S$ & $1989$ & $1968$ \\
    & $2S$ & $2705$ & - \\[2ex]
$D^*_s$ & $1S$ & $2115$ & $2112$ \\
      & $2S$ & $2769$ & - \\[2ex]
$\eta_c$ & $1S$ & $2989$ & $2981$ \\
    & $2S$ & $3627$ & - \\[2ex]
$J/\psi$ & $1S$ & $3097$ & $3097$ \\
 $\psi$   & $2S$ & $3685$ & - 
\end{tabular}
\end{ruledtabular}
\end{table}

\begin{figure}[ht]
\epsfxsize=3.4in \epsfbox{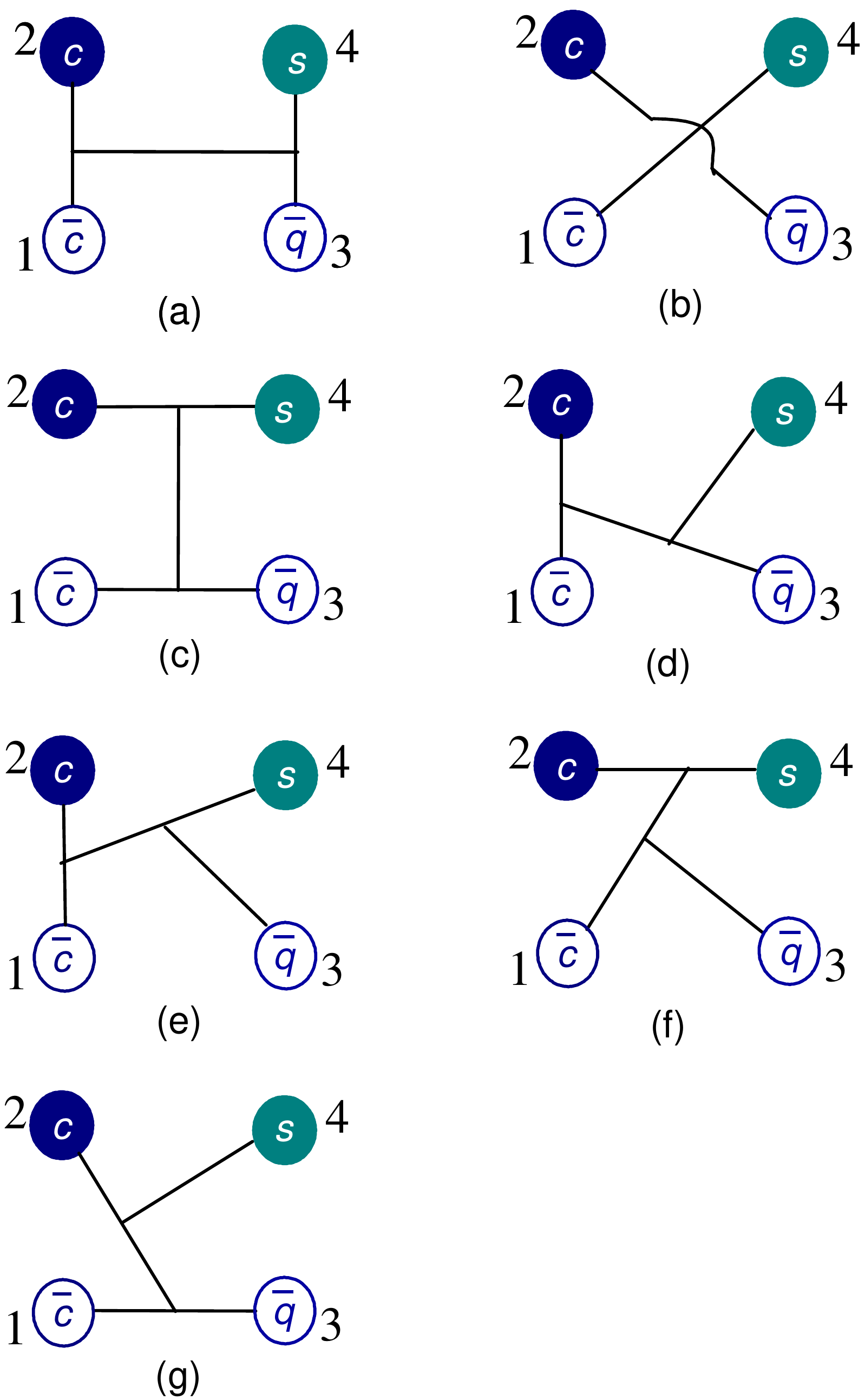}
\caption{Seven types of configurations in $c\bar{c}s\bar{q}$ $(q=u,\,d)$ tetraquarks. Panel $(a)$ and $(b)$ are meson-meson structures, panel $(c)$ is diquark-antidiquark one and the other K-type structures are from panel $(d)$ to $(g)$.} \label{QQqq}
\end{figure}

Figure~\ref{QQqq} shows seven kinds of configurations for the $c\bar{c}s\bar{q}$ tetraquark system. In particular, Fig.~\ref{QQqq}(a) and Fig.~\ref{QQqq}(b) are meson-meson structures, Fig.~\ref{QQqq}(c) is the diquark-antidiquark one, and the other K-type configurations are from panels (d) to (g). All of them, and their couplings, are considered in our investigation. However, for the purpose of solving a manageable 4-body problem, the K-type configurations are sometimes restricted. It is important to note herein that just one configuration would be enough for the calculation, if all radial and orbital excited states were taken into account; however, this is obviously much less efficient and thus an economic way is to combine the different configurations in the ground state to perform the calculation.

The multiquark system's wave function at the quark level is an internal product of color, spin, flavor and space terms. Concerning the color degree-of-freedom, the colorless wave function of a 4-quark system in meson-meson configuration, as illustrated in Fig.~\ref{QQqq}(a) and Fig.~\ref{QQqq}(b), can be obtained by either two coupled color-singlet clusters, $1\otimes 1$:
\begin{align}
\label{Color1}
\chi^c_1 &= \frac{1}{3}(\bar{r}r+\bar{g}g+\bar{b}b)\times (\bar{r}r+\bar{g}g+\bar{b}b) \,,
\end{align}
or two coupled color-octet clusters, $8\otimes 8$:
\begin{align}
\label{Color2}
\chi^c_2 &= \frac{\sqrt{2}}{12}(3\bar{b}r\bar{r}b+3\bar{g}r\bar{r}g+3\bar{b}g\bar{g}b+3\bar{g}b\bar{b}g+3\bar{r}g\bar{g}r
\nonumber\\
&+3\bar{r}b\bar{b}r+2\bar{r}r\bar{r}r+2\bar{g}g\bar{g}g+2\bar{b}b\bar{b}b-\bar{r}r\bar{g}g
\nonumber\\
&-\bar{g}g\bar{r}r-\bar{b}b\bar{g}g-\bar{b}b\bar{r}r-\bar{g}g\bar{b}b-\bar{r}r\bar{b}b) \,.
\end{align}
These two color states are the so-called color-singlet and hidden-color channels, respectively.

The color wave functions associated to the diquark-antidiquark structure shown in Fig.~\ref{QQqq}(c) are the coupled color triplet-antitriplet clusters, $3\otimes \bar{3}$:
\begin{align}
\label{Color3}
\chi^c_3 &= \frac{\sqrt{3}}{6}(\bar{r}r\bar{g}g-\bar{g}r\bar{r}g+\bar{g}g\bar{r}r-\bar{r}g\bar{g}r+\bar{r}r\bar{b}b
\nonumber\\
&-\bar{b}r\bar{r}b+\bar{b}b\bar{r}r-\bar{r}b\bar{b}r+\bar{g}g\bar{b}b-\bar{b}g\bar{g}b
\nonumber\\
&+\bar{b}b\bar{g}g-\bar{g}b\bar{b}g) \,,
\end{align}
and the coupled color sextet-antisextet clusters, $6\otimes \bar{6}$:
\begin{align}
\label{Color4}
\chi^c_4 &= \frac{\sqrt{6}}{12}(2\bar{r}r\bar{r}r+2\bar{g}g\bar{g}g+2\bar{b}b\bar{b}b+\bar{r}r\bar{g}g+\bar{g}r\bar{r}g
\nonumber\\
&+\bar{g}g\bar{r}r+\bar{r}g\bar{g}r+\bar{r}r\bar{b}b+\bar{b}r\bar{r}b+\bar{b}b\bar{r}r
\nonumber\\
&+\bar{r}b\bar{b}r+\bar{g}g\bar{b}b+\bar{b}g\bar{g}b+\bar{b}b\bar{g}g+\bar{g}b\bar{b}g) \,.
\end{align}

Meanwhile, the colorless wave functions of the K-type structures shown in Fig.~\ref{QQqq}(d) to (g) are 
\begin{align}
\label{Color5}
\chi^c_5 &= \frac{1}{6\sqrt{2}}(\bar{r}r\bar{r}r+\bar{g}g\bar{g}g-2\bar{b}b\bar{b}b)+
\nonumber\\
&\frac{1}{2\sqrt{2}}(\bar{r}b\bar{b}r+\bar{r}g\bar{g}r+\bar{g}b\bar{b}g+\bar{g}r\bar{r}g+\bar{b}g\bar{g}b+\bar{b}r\bar{r}b)-
\nonumber\\
&\frac{1}{3\sqrt{2}}(\bar{g}g\bar{r}r+\bar{r}r\bar{g}g)+\frac{1}{6\sqrt{2}}(\bar{b}b\bar{r}r+\bar{b}b\bar{g}g+\bar{r}r\bar{b}b+\bar{g}g\bar{b}b) \,,
\end{align}
\begin{align}
\label{Color6}
\chi^c_6 &= \chi^c_1 \,,
\end{align}
\begin{align}
\label{Color7}
\chi^c_7 &= \chi^c_1 \,,
\end{align}
\begin{align}
\label{Color8}
\chi^c_8 &= \frac{1}{4}(1-\frac{1}{\sqrt{6}})\bar{r}r\bar{g}g-\frac{1}{4}(1+\frac{1}{\sqrt{6}})\bar{g}g\bar{g}g-\frac{1}{4\sqrt{3}}\bar{r}g\bar{g}r+
\nonumber\\
&\frac{1}{2\sqrt{2}}(\bar{r}b\bar{b}r+\bar{g}b\bar{b}g+\bar{b}g\bar{g}b+\bar{g}r\bar{r}g+\bar{b}r\bar{r}b)+
\nonumber\\
&\frac{1}{2\sqrt{6}}(\bar{r}r\bar{b}b-\bar{g}g\bar{b}b+\bar{b}b\bar{g}g+\bar{g}g\bar{r}r-\bar{b}b\bar{r}r) \,,
\end{align}
\begin{align}
\label{Color9}
\chi^c_9 &= \frac{1}{2\sqrt{6}}(\bar{r}b\bar{b}r+\bar{r}r\bar{b}b+\bar{g}b\bar{b}g+\bar{g}g\bar{b}b+\bar{r}g\bar{g}r+\bar{r}r\bar{g}g+
\nonumber\\
&\bar{b}b\bar{g}g+\bar{b}g\bar{g}b+\bar{g}g\bar{r}r+\bar{g}r\bar{r}g+\bar{b}b\bar{r}r+\bar{b}r\bar{r}b)+
\nonumber\\
&\frac{1}{\sqrt{6}}(\bar{r}r\bar{r}r+\bar{g}g\bar{g}g+\bar{b}b\bar{b}b) \,,
\end{align}
\begin{align}
\label{Color10}
\chi^c_{10} &= \frac{1}{2\sqrt{3}}(\bar{r}b\bar{b}r-\bar{r}r\bar{b}b+\bar{g}b\bar{b}g-\bar{g}g\bar{b}b+\bar{r}g\bar{g}r-\bar{r}r\bar{g}g-
\nonumber\\
&\bar{b}b\bar{g}g+\bar{b}g\bar{g}b-\bar{g}g\bar{r}r+\bar{g}r\bar{r}g-\bar{b}b\bar{r}r+\bar{b}r\bar{r}b) \,,
\end{align}
\begin{align}
\label{Color11}
\chi^c_{11} &= \chi^c_9 \,,
\end{align}
\begin{align}
\label{Color12}
\chi^c_{12} &= -\chi^c_{10} \,.
\end{align}

As for the flavor degree-of-freedom, since the quark content of the investigated tetraquark system is $c\bar{c}s\bar{q}$ $(q=u,\,d)$, only $I=1/2$ sector is discussed. The flavor wave-function is denoted as $\chi^{f}_{I, M_I}$, where the third component of the isospin, $M_I$, is fixed to be equal to $I$ for simplicity, since the Hamiltonian does not have a flavor-dependent interaction which can distinguish the third component of the isospin quantum number.

We are going to considered $S$-wave ground states with spin ranging from $S=0$ to $2$. Therefore, the spin wave functions, $\chi^{\sigma_i}_{S, M_S}$, are given by ($M_S$ can be set to be equal to $S$ without loss of generality):
\begin{align}
\label{SWF1}
\chi_{0,0}^{\sigma_{u_1}}(4) &= \chi^\sigma_{00}\chi^\sigma_{00} \,, \\
\chi_{0,0}^{\sigma_{u_2}}(4) &= \frac{1}{\sqrt{3}}(\chi^\sigma_{11}\chi^\sigma_{1,-1}-\chi^\sigma_{10}\chi^\sigma_{10}+\chi^\sigma_{1,-1}\chi^\sigma_{11}) \,, \\
\chi_{0,0}^{\sigma_{u_3}}(4) &= \frac{1}{\sqrt{2}}\big((\sqrt{\frac{2}{3}}\chi^\sigma_{11}\chi^\sigma_{\frac{1}{2}, -\frac{1}{2}}-\sqrt{\frac{1}{3}}\chi^\sigma_{10}\chi^\sigma_{\frac{1}{2}, \frac{1}{2}})\chi^\sigma_{\frac{1}{2}, -\frac{1}{2}} \nonumber \\ 
&-(\sqrt{\frac{1}{3}}\chi^\sigma_{10}\chi^\sigma_{\frac{1}{2}, -\frac{1}{2}}-\sqrt{\frac{2}{3}}\chi^\sigma_{1, -1}\chi^\sigma_{\frac{1}{2}, \frac{1}{2}})\chi^\sigma_{\frac{1}{2}, \frac{1}{2}}\big) \,, \\
\chi_{0,0}^{\sigma_{u_4}}(4) &= \frac{1}{\sqrt{2}}(\chi^\sigma_{00}\chi^\sigma_{\frac{1}{2}, \frac{1}{2}}\chi^\sigma_{\frac{1}{2}, -\frac{1}{2}}-\chi^\sigma_{00}\chi^\sigma_{\frac{1}{2}, -\frac{1}{2}}\chi^\sigma_{\frac{1}{2}, \frac{1}{2}}) \,,
\end{align}
\begin{align}
\chi_{1,1}^{\sigma_{w_1}}(4) &= \chi^\sigma_{00}\chi^\sigma_{11} \,, \\ 
\chi_{1,1}^{\sigma_{w_2}}(4) &= \chi^\sigma_{11}\chi^\sigma_{00} \,, \\
\chi_{1,1}^{\sigma_{w_3}}(4) &= \frac{1}{\sqrt{2}} (\chi^\sigma_{11} \chi^\sigma_{10}-\chi^\sigma_{10} \chi^\sigma_{11}) \,, \\
\chi_{1,1}^{\sigma_{w_4}}(4) &= \sqrt{\frac{3}{4}}\chi^\sigma_{11}\chi^\sigma_{\frac{1}{2}, \frac{1}{2}}\chi^\sigma_{\frac{1}{2}, -\frac{1}{2}}-\sqrt{\frac{1}{12}}\chi^\sigma_{11}\chi^\sigma_{\frac{1}{2}, -\frac{1}{2}}\chi^\sigma_{\frac{1}{2}, \frac{1}{2}} \nonumber \\ 
&-\sqrt{\frac{1}{6}}\chi^\sigma_{10}\chi^\sigma_{\frac{1}{2}, \frac{1}{2}}\chi^\sigma_{\frac{1}{2}, \frac{1}{2}} \,, \\
\chi_{1,1}^{\sigma_{w_5}}(4) &= (\sqrt{\frac{2}{3}}\chi^\sigma_{11}\chi^\sigma_{\frac{1}{2}, -\frac{1}{2}}-\sqrt{\frac{1}{3}}\chi^\sigma_{10}\chi^\sigma_{\frac{1}{2}, \frac{1}{2}})\chi^\sigma_{\frac{1}{2}, \frac{1}{2}} \,, \\
\chi_{1,1}^{\sigma_{w_6}}(4) &= \chi^\sigma_{00}\chi^\sigma_{\frac{1}{2}, \frac{1}{2}}\chi^\sigma_{\frac{1}{2}, \frac{1}{2}} \,, \\
\label{SWF2}
\chi_{2,2}^{\sigma_{1}}(4) &= \chi^\sigma_{11}\chi^\sigma_{11} \,.
\end{align}
The superscripts $u_1,\ldots,u_4$ and $w_1,\ldots,w_6$ determine the spin wave function for each configuration of the $c\bar{c}s\bar{q}$ tetraquark system, their specific values are shown in Table~\ref{SpinIndex}. Furthermore, the expressions above are obtained by considering the coupling of two sub-clusters whose spin wave functions are given by trivial SU(2) algebra, and the necessary basis reads as
\begin{align}
\label{Spin}
\chi^\sigma_{11} &= \chi^\sigma_{\frac{1}{2}, \frac{1}{2}} \chi^\sigma_{\frac{1}{2}, \frac{1}{2}} \,, \\
\chi^\sigma_{1,-1} &= \chi^\sigma_{\frac{1}{2}, -\frac{1}{2}} \chi^\sigma_{\frac{1}{2}, -\frac{1}{2}} \,, \\
\chi^\sigma_{10} &= \frac{1}{\sqrt{2}}(\chi^\sigma_{\frac{1}{2}, \frac{1}{2}} \chi^\sigma_{\frac{1}{2}, -\frac{1}{2}}+\chi^\sigma_{\frac{1}{2}, -\frac{1}{2}} \chi^\sigma_{\frac{1}{2}, \frac{1}{2}}) \,, \\
\chi^\sigma_{00} &= \frac{1}{\sqrt{2}}(\chi^\sigma_{\frac{1}{2}, \frac{1}{2}} \chi^\sigma_{\frac{1}{2}, -\frac{1}{2}}-\chi^\sigma_{\frac{1}{2}, -\frac{1}{2}} \chi^\sigma_{\frac{1}{2}, \frac{1}{2}}) \,, 
\end{align}

\begin{table}[!t]
\caption{\label{SpinIndex} The values of the superscripts $u_1,\ldots,u_4$ and $w_1,\ldots,w_6$ that determine the spin wave function for each configuration of the $c\bar{c}s\bar{q}$ tetraquark system.}
\begin{ruledtabular}
\begin{tabular}{lcccccc}
& Di-meson & Diquark-antidiquark & $K_1$ & $K_2$ & $K_3$ & $K_4$ \\
\hline
$u_1$ & 1 & 3 & & & & \\
$u_2$ & 2 & 4 & & & & \\
$u_3$ &   &   & 5 & 7 &  9 & 11 \\
$u_4$ &   &   & 6 & 8 & 10 & 12 \\[2ex]
$w_1$ & 1 & 4 & & & & \\
$w_2$ & 2 & 5 & & & & \\
$w_3$ & 3 & 6 & & & & \\
$w_4$ &   &   & 7 & 10 & 13 & 16 \\
$w_5$ &   &   & 8 & 11 & 14 & 17 \\
$w_6$ &   &   & 9 & 12 & 15 & 18
\end{tabular}
\end{ruledtabular}
\end{table}

Among the different methods to solve the Schr\"odinger-like 4-body bound state equation, we use the Rayleigh-Ritz variational principle which is one of the most extended tools to solve eigenvalue problems because its simplicity and flexibility. Moreover, we use the complex-range method and thus the spatial wave function is written as follows:
\begin{equation}
\label{eq:WFexp}
\psi_{LM_L}(\theta)= \left[ \left[ \phi_{n_1l_1}(\vec{\rho}e^{i\theta}\,) \phi_{n_2l_2}(\vec{\lambda}e^{i\theta}\,)\right]_{l} \phi_{n_3l_3}(\vec{R}e^{i\theta}\,) \right]_{L M_L} \,,
\end{equation}
where the internal Jacobi coordinates are defined as
\begin{align}
\vec{\rho} &= \vec{x}_1-\vec{x}_{2(4)} \,, \\
\vec{\lambda} &= \vec{x}_3 - \vec{x}_{4(2)} \,, \\
\vec{R} &= \frac{m_1 \vec{x}_1 + m_{2(4)} \vec{x}_{2(4)}}{m_1+m_{2(4)}}- \frac{m_3 \vec{x}_3 + m_{4(2)} \vec{x}_{4(2)}}{m_3+m_{4(2)}} \,,
\end{align}
for the meson-meson configurations of Fig.~\ref{QQqq}(a) and \ref{QQqq}(b), where the numbers in parentheses are those corresponding to Fig.~\ref{QQqq}(b); and as
\begin{align}
\vec{\rho} &= \vec{x}_1-\vec{x}_3 \,, \\
\vec{\lambda} &= \vec{x}_2 - \vec{x}_4 \,, \\
\vec{R} &= \frac{m_1 \vec{x}_1 + m_3 \vec{x}_3}{m_1+m_3}- \frac{m_2 \vec{x}_2 + m_4 \vec{x}_4}{m_2+m_4} \,,
\end{align}
for the diquark-antdiquark structure of Fig.~\ref{QQqq}(c). The remaining K-type configurations shown in Fig.~\ref{QQqq}(d) to \ref{QQqq}(g) are ($i, j, k, l$ take values according to the panels (d) to (g) of Fig.~\ref{QQqq}):
\begin{align}
\vec{\rho} &= \vec{x}_i-\vec{x}_j \,, \\
\vec{\lambda} &= \vec{x}_k- \frac{m_i \vec{x}_i + m_j \vec{x}_j}{m_i+m_j} \,, \\
\vec{R} &= \vec{x}_l- \frac{m_i \vec{x}_i + m_j \vec{x}_j+m_k \vec{x}_k}{m_i+m_j+m_k} \,.
\end{align}
It becomes obvious now that the center-of-mass kinetic term, $T_{\text{CM}}$, can be completely eliminated for a non-relativistic system defined in any of the above sets of relative coordinates.

A crucial aspect of the Rayleigh-Ritz variational method is the basis expansion of the trial wave function. We are going to use the Gaussian expansion method (GEM)~\cite{Hiyama:2003cu} in which each relative coordinate is expanded in terms of Gaussian basis functions whose sizes are taken in geometric progression. This method has proven to be very efficient on solving the bound-state problem of multiquark systems~\cite{Yang:2015bmv, Yang:2018oqd, gy:2020dcp, gy:2020dht, gy:2020dhts, Yang:2021izl, YangSym2020} and the details on how the geometric progression is fixed can be found in, \emph{e.g}, Ref.~\cite{Yang:2015bmv}. Therefore, the form of the orbital wave functions, $\phi$'s, in Eq.~\eqref{eq:WFexp} is 
\begin{align}
&
\phi_{nlm}(\vec{r}e^{i\theta}\,) = N_{nl} (re^{i\theta})^{l} e^{-\nu_{n} (re^{i\theta})^2} Y_{lm}(\hat{r}) \,.
\end{align}
Since only $S$-wave states of charm(bottom)-strange tetraquarks are investigated in this work, no laborious Racah algebra is needed while computing matrix elements. In this case, the value of the spherical harmonic function is just a constant, \emph{viz.} $Y_{00}=\sqrt{1/4\pi}$.

Finally, the complete wave-function that fulfills the Pauli principle is written as
\begin{equation}
\label{TPs}
\Psi_{JM_J,I,i,j,k}(\theta)={\cal A} \left[ \left[ \psi_{L}(\theta) \chi^{\sigma_i}_{S}(4) \right]_{JM_J} \chi^{f_j}_I \chi^{c}_k \right] \,,
\end{equation}
where $\cal{A}$ is the antisymmetry operator of $c\bar{c}s\bar{q}$ tetraquark system and it just reads $\cal{A}$ = 1, since each of the four particles are nonidentical.


\section{Results}
\label{sec:results}

\begin{table}[!t]
\caption{\label{GDD1} All possible channels for $J^P=0^+$ $c\bar{c}s\bar{q}$ tetraquark system. The second column shows the necessary basis combination in spin ($\chi_J^{\sigma_i}$), flavor ($\chi_I^{f_j}$) and color ($\chi_k^c$) degrees of freedom. Particularly, the flavor index ($j$) 1 is of $\bar{c}c\bar{q}s$ and 2 is of $\bar{c}s\bar{q}c$, respectively. The superscript 1 and 8 stands for the color-singlet and hidden-color configurations of physical channels.}
\begin{ruledtabular}
\begin{tabular}{ccc}
~~Index & $\chi_J^{\sigma_i}$;~$\chi_I^{f_j}$;~$\chi_k^c$ & Channel~~\\
              &$[i; ~j; ~k]$&  \\[2ex]
1  & $[1; ~1; ~1]$   & $(\eta_c K)^1$ \\
2 & $[2; ~1; ~1]$  & $(J/\psi K^*)^1$ \\
3 & $[1; ~2; ~1]$  & $(DD_s)^1$ \\
4 & $[2; ~2; ~1]$  & $(D^* D^*_s)^1$ \\
5 & $[1; ~1; ~2]$  & $(\eta_c K)^8$ \\
6 & $[2; ~1; ~2]$  & $(J/\psi K^*)^8$ \\
7 & $[1; ~2; ~2]$  & $(DD_s)^8$ \\
8 & $[2; ~2; ~2]$   & $(D^* D^*_s)^8$ \\
9 & $[3; ~1; ~3]$  & $(cs)_3 (\bar{c}\bar{q})_{\bar{3}}$ \\
10 & $[4; ~1; ~3]$  & $(cs)^*_3 (\bar{c}\bar{q})^*_{\bar{3}}$ \\
11 & $[3; ~1; ~4]$  & $(cs)_6 (\bar{c}\bar{q})_{\bar{6}}$ \\
12 & $[4; ~1; ~4]$  & $(cs)^*_6 (\bar{c}\bar{q})^*_{\bar{6}}$ \\
13 & $[5; ~1; ~5]$  & $K_1$ \\
14 & $[6; ~1; ~5]$  & $K_1$ \\
15 & $[5; ~1; ~6]$   & $K_1$ \\
16 & $[6; ~1; ~6]$  & $K_1$ \\
17 & $[7; ~1; ~7]$  & $K_2$ \\
18 & $[8; ~1; ~7]$  & $K_2$\\
19 & $[7; ~1; ~8]$  & $K_2$ \\
20 & $[8; ~1; ~8]$  & $K_2$ \\
21 & $[9; ~1; ~9]$  & $K_3$ \\
22 & $[10; ~1; ~9]$  & $K_3$ \\
23 & $[9; ~1; ~10]$  & $K_3$ \\
24 & $[10; ~1; ~10]$  & $K_3$ \\
25 & $[11; ~1; ~11]$   & $K_4$ \\
26 & $[12; ~1; ~11]$  & $K_4$ \\
27 & $[11; ~1; ~12]$  & $K_4$ \\
28 & $[12; ~1; ~12]$  & $K_4$\\
\end{tabular}
\end{ruledtabular}
\end{table}

\begin{table*}[!t]
\caption{\label{GDD2} All possible channels for $J^P=1^+$ $c\bar{c}s\bar{q}$ tetraquark system. The second and fifth columns show the necessary basis combination in spin ($\chi_J^{\sigma_i}$), flavor ($\chi_I^{f_j}$) and color ($\chi_k^c$) degrees of freedom. Particularly, the flavor indices ($j$) 1 is of $\bar{c}c\bar{q}s$ and 2 is of $\bar{c}s\bar{q}c$, respectively. The superscript 1 and 8 stands for the color-singlet and hidden-color configurations of physical channels.}
\begin{ruledtabular}
\begin{tabular}{cccccc}
Index~ & $\chi_J^{\sigma_i}$;~$\chi_I^{f_j}$;~$\chi_k^c$ & Channel~~ & Index~~ & $\chi_J^{\sigma_i}$;~$\chi_I^{f_j}$;~$\chi_k^c$ & Channel~~ \\
&$[i; ~j; ~k]$ & & &$[i; ~j; ~k]$&  \\[2ex]
 1  & $[1; ~1; ~1]$   & $(\eta_c K^*)^1$ & 19 & $[7; ~1; ~5]$   & $K_1$ \\
 2 & $[2; ~1; ~1]$ & $(J/\psi K)^1$ & 20 & $[8; ~1; ~5]$  & $K_1$ \\
 3 & $[3; ~1; ~1]$   & $(J/\psi K^*)^1$ & 21  & $[9; ~1; ~5]$   & $K_1$ \\
 4 & $[1; ~2; ~1]$ & $(DD^*_s)^1$ & 22 & $[7; ~1; ~6]$   & $K_1$   \\
 5  & $[2; ~2; ~1]$ & $(D^* D_s)^1$  & 23 & $[8; ~1; ~6]$   & $K_1$ \\
 6 & $[3; ~2; ~1]$ & $(D^* D^*_s)^1$ & 24  & $[9; ~1; ~6]$   & $K_1$ \\
 7  & $[1; ~1; ~2]$   & $(\eta_c K^*)^8$ & 25 & $[10; ~1; ~7]$   & $K_2$ \\
 8 & $[2; ~1; ~2]$ & $(J/\psi K)^8$ & 26 & $[11; ~1; ~7]$  & $K_2$ \\
 9 & $[3; ~1; ~2]$   & $(J/\psi K^*)^8$ & 27  & $[12; ~1; ~7]$   & $K_2$ \\
 10 & $[1; ~2; ~2]$ & $(DD^*_s)^8$ & 28 & $[10; ~1; ~8]$   & $K_2$   \\
 11  & $[2; ~2; ~2]$ & $(D^* D_s)^8$  & 29 & $[11; ~1; ~8]$   & $K_2$ \\
 12 & $[3; ~2; ~2]$ & $(D^* D^*_s)^8$ & 30  & $[12; ~1; ~8]$   & $K_2$ \\
 13  & $[4; ~1; ~3]$ & $(cs)_3 (\bar{c}\bar{q})^*_{\bar{3}}$ & 31 & $[13; ~1; ~9]$& $K_3$ \\
 14 & $[5; ~1; ~3]$ & $(cs)^*_3 (\bar{c}\bar{q})_{\bar{3}}$  & 32 & $[14; ~1; ~9]$ & $K_3$ \\
 15 & $[6; ~1; ~3]$ & $(cs)^*_3 (\bar{c}\bar{q})^*_{\bar{3}}$ & 33 & $[15; ~1; ~9]$ & $K_3$\\
 16 & $[4; ~1; ~4]$ & $(cs)_6 (\bar{c}\bar{q})^*_{\bar{6}}$  & 34 & $[13; ~1; ~10]$   & $K_3$   \\
 17  & $[5; ~1; ~4]$ & $(cs)^*_6 (\bar{c}\bar{q})_{\bar{6}}$   & 35 & $[14; ~1; ~10]$   & $K_3$ \\
 18 & $[6; ~1; ~4]$ & $(cs)^*_6 (\bar{c}\bar{q})^*_{\bar{6}}$  & 36  & $[15; ~1; ~10]$   & $K_3$ \\
  &  &  & 37  & $[16; ~1; ~11]$   & $K_4$ \\
  &  &  & 38  & $[17; ~1; ~11]$   & $K_4$ \\
  &  &  & 39  & $[18; ~1; ~11]$   & $K_4$ \\
  &  &  & 40  & $[16; ~1; ~12]$   & $K_4$ \\
  &  &  & 41  & $[17; ~1; ~12]$   & $K_4$ \\
  &  &  & 42  & $[18; ~1; ~12]$   & $K_4$ \\
\end{tabular}
\end{ruledtabular}
\end{table*}

\begin{table}[!t]
\caption{\label{GDD3} All possible channels for $J^P=2^+$ $c\bar{c}s\bar{q}$ tetraquark system. The second column shows the necessary basis combination in spin ($\chi_J^{\sigma_i}$), flavor ($\chi_I^{f_j}$) and color ($\chi_k^c$) degrees of freedom. Particularly, the flavor indices ($j$) 1 is of $\bar{c}c\bar{q}s$ and 2 is of $\bar{c}s\bar{q}c$, respectively. The superscript 1 and 8 stands for the color-singlet and hidden-color configurations of physical channels.}
\begin{ruledtabular}
\begin{tabular}{ccc}
~~Index & $\chi_J^{\sigma_i}$;~$\chi_I^{f_j}$;~$\chi_k^c$ & Channel~~\\
              &$[i; ~j; ~k]$&  \\[2ex]
1 & $[1; ~1; ~1]$  & $(J/\psi K^*)^1$ \\
2 & $[1; ~2; ~1]$  & $(D^* D^*_s)^1$ \\
3 & $[1; ~1; ~2]$  & $(J/\psi K^*)^8$ \\
4 & $[1; ~2; ~2]$   & $(D^* D^*_s)^8$ \\
5 & $[1; ~1; ~3]$  & $(cs)^*_3 (\bar{c}\bar{q})^*_{\bar{3}}$ \\
6 & $[1; ~1; ~4]$  & $(cs)^*_6 (\bar{c}\bar{q})^*_{\bar{6}}$ \\
7 & $[1; ~1; ~5]$  & $K_1$ \\
8 & $[1; ~1; ~6]$  & $K_1$ \\
9 & $[1; ~1; ~7]$  & $K_2$ \\
10 & $[1; ~1; ~8]$  & $K_2$ \\
11 & $[1; ~1; ~9]$  & $K_3$ \\
12 & $[1; ~1; ~10]$  & $K_3$ \\
13 & $[1; ~1; ~11]$   & $K_4$ \\
14 & $[1; ~1; ~12]$  & $K_4$\\
\end{tabular}
\end{ruledtabular}
\end{table}

In the present calculation, we investigate all possible $S$-wave hidden-charm tetraquarks with strangeness by taking into account di-meson, diquark-antidiquark and K-type configurations. In our approach, a $c\bar{c}s\bar{q}$ tetraquark state has positive parity assuming that the angular momenta $l_1$, $l_2$ and $l_3$ in Eq.~\eqref{eq:WFexp} are all equal to zero. Accordingly, the total angular momentum, $J$, coincides with the total spin, $S$, and can take values $0$, $1$ and $2$. Besides, the value of isospin, $I$, can only be $1/2$ considering the quark content of the $c\bar{c}s\bar{q}$ system. 

Tables~\ref{GDD1},~\ref{GDD2} and~\ref{GDD3} list the allowed meson-meson, diquark-antidiquark and K-type channels; they are indexed in the first column, particular combinations of spin ($\chi_J^{\sigma_i}$), flavor ($\chi_I^{f_j}$) and color ($\chi_k^c$) wave functions are shown in the second column, and the last column reflects the specific physical channel.

Let us proceed now to describe in detail our theoretical findings for each $J^P=0^+$, $1^+$ and $2^+$ sector of $c\bar{c}s\bar{q}$ tetraquarks. Three subsections are presented in the following and, in order to explore the detailed nature of the found states, three kinds of calculations are performed, \emph{i.e.} the $(c\bar{c})(s\bar{q})$ di-meson configuration along with diquark-antidiquark and K-type ones in a coupled-channels study, the $(c\bar{q})(s\bar{c})$ di-meson configuration coupled with the other two types of exotic structures, and a complete coupled-channels investigation.


\subsection{The $\mathbf{J^P=0^+}$ $\mathbf{c\bar{c}s\bar{q}}$ tetraquark system}

\begin{table}[!t]
\caption{\label{GresultCC1} Lowest-lying $c\bar{c}s\bar{q}$ tetraquark states with $J^P=0^+$ calculated within the real range formulation of the chiral quark model.
The allowed meson-meson, diquark-antidiquark and K-type configurations are listed in the first column; when possible, the experimental value of the non-interacting meson-meson threshold is labeled in parentheses. Each channel is assigned an index in the 2nd column. The theoretical mass obatined in each channel is shown in the 3rd column and the coupled result for each kind of configuration is presented in the last column.
When a complete coupled-channels calculation is performed, last row of the table indicates the lowest-lying mass.
(unit: MeV).}
\begin{ruledtabular}
\begin{tabular}{lccc}
~~Channel   & Index & $M$ & Mixed~~ \\[2ex]
$(\eta_c K)^1 (3475)$          & 1   & $3470$ & \\
$(J/\psi K^*)^1 (3989)$  & 2    & $4004$ & \\
$(D D_s)^1 (3838)$      & 3     & $3886$ & \\
$(D^* D^*_s)^1 (4119)$  & 4     & $4132$ & $3470$ \\[2ex]
$(\eta_c K)^8$          & 5   & $4422$ & \\
$(J/\psi K^*)^8$  & 6     & $4416$ &  \\
$(D D_s)^8$      & 7     & $4398$ & \\
$(D^* D^*_s)^8$  & 8     & $4333$ & $4177$ \\[2ex]
$(cs)_3 (\bar{c}\bar{q})_{\bar{3}}$      & 9     & $4320$ & \\
$(cs)^*_3 (\bar{c}\bar{q})^*_{\bar{3}}$      & 10    & $4334$ & \\
$(cs)_6 (\bar{c}\bar{q})_{\bar{6}}$  & 11     & $4349$ & \\
$(cs)^*_6 (\bar{c}\bar{q})^*_{\bar{6}}$  & 12     & $4224$ & $4122$ \\[2ex]
$K_1$  & 13     & $4407$ & \\
  & 14     & $4420$ & \\
  & 15     & $4296$ & \\
  & 16     & $4080$ & $4075$ \\[2ex]
$K_2$  & 17     & $4283$ & \\
  & 18    & $4062$ & \\
  & 19     & $4407$ & \\
  & 20     & $4419$ & $4057$ \\[2ex]
$K_3$  & 21    & $4217$ & \\
  & 22     & $4353$ & \\
  & 23     & $4320$ & \\
  & 24     & $4319$ & $4114$ \\[2ex]
$K_4$  & 25     & $4241$ & \\
    & 26     & $4372$ & \\
    & 27    & $4326$ & \\
      & 28     & $4324$ & $4117$ \\[2ex]
\multicolumn{3}{c}{Complete coupled-channels:} & $3470$ \\
\end{tabular}
\end{ruledtabular}
\end{table}

\begin{figure}[ht]
\includegraphics[width=0.49\textwidth, trim={2.3cm 2.0cm 2.0cm 1.0cm}]{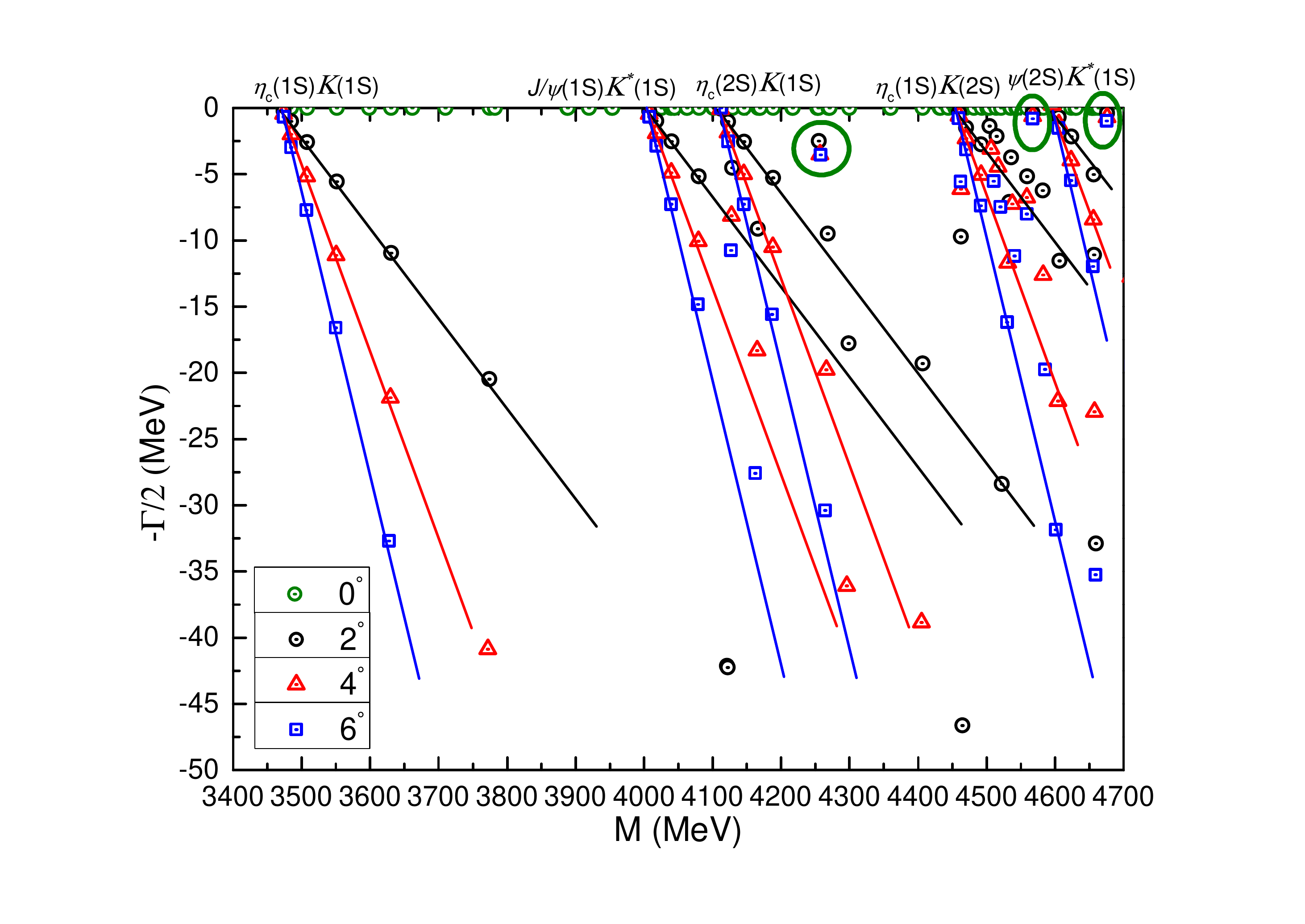}
\caption{\label{PP1} The coupled-channels calculation of the $c\bar{c}s\bar{q}$ tetraquark system with $J^P=0^+$ quantum numbers. Particularly, the $(c\bar{q})(s\bar{c})$ dimeson channels are excluded. We use the complex-scaling method of the chiral quark model varying $\theta$ from $0^\circ$ to $6^\circ$.}
\end{figure}

\begin{figure}[!t]
\includegraphics[clip, trim={3.0cm 2.0cm 3.0cm 1.0cm}, width=0.45\textwidth]{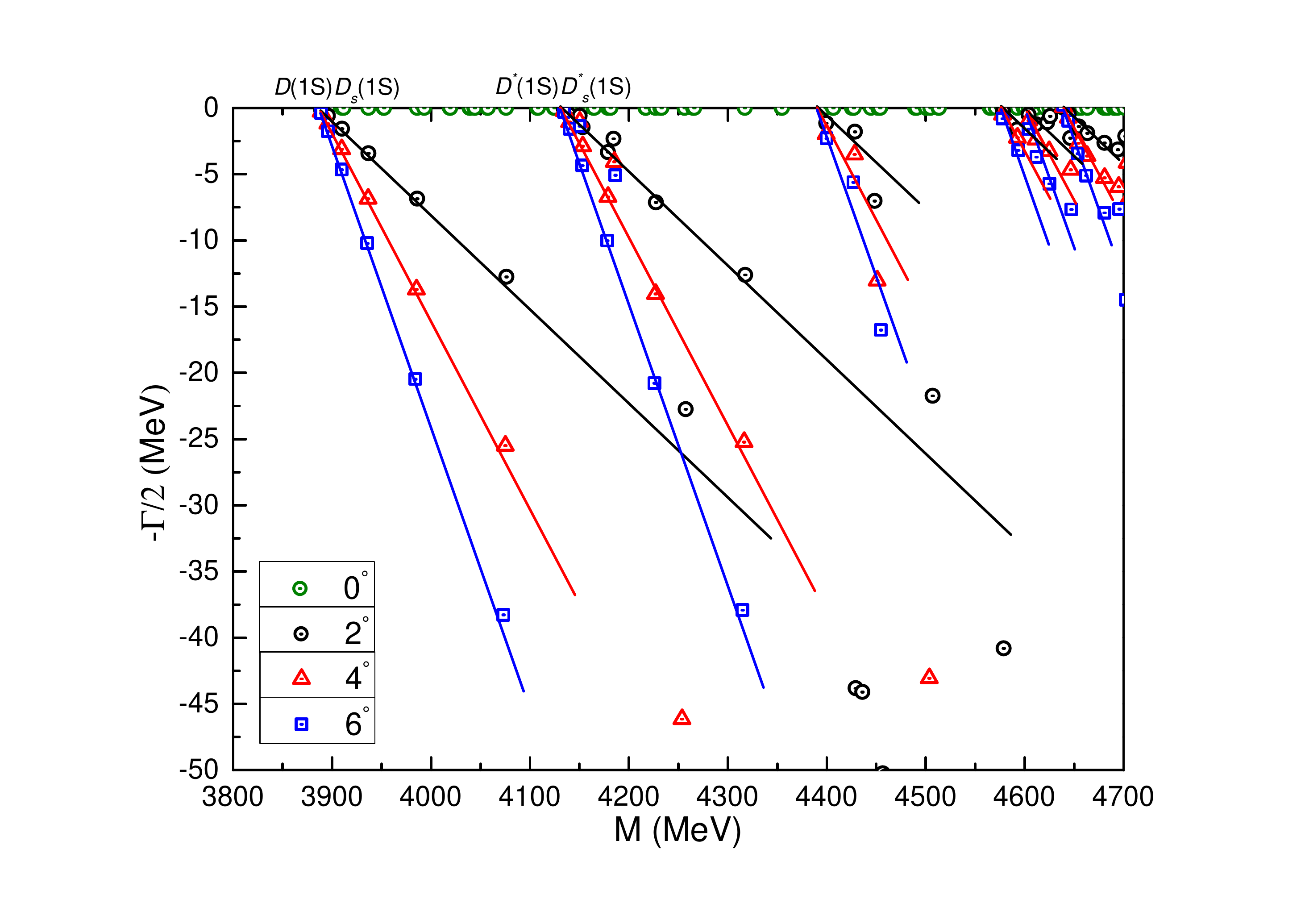} \\
\includegraphics[clip, trim={3.0cm 2.0cm 3.0cm 1.0cm}, width=0.45\textwidth]{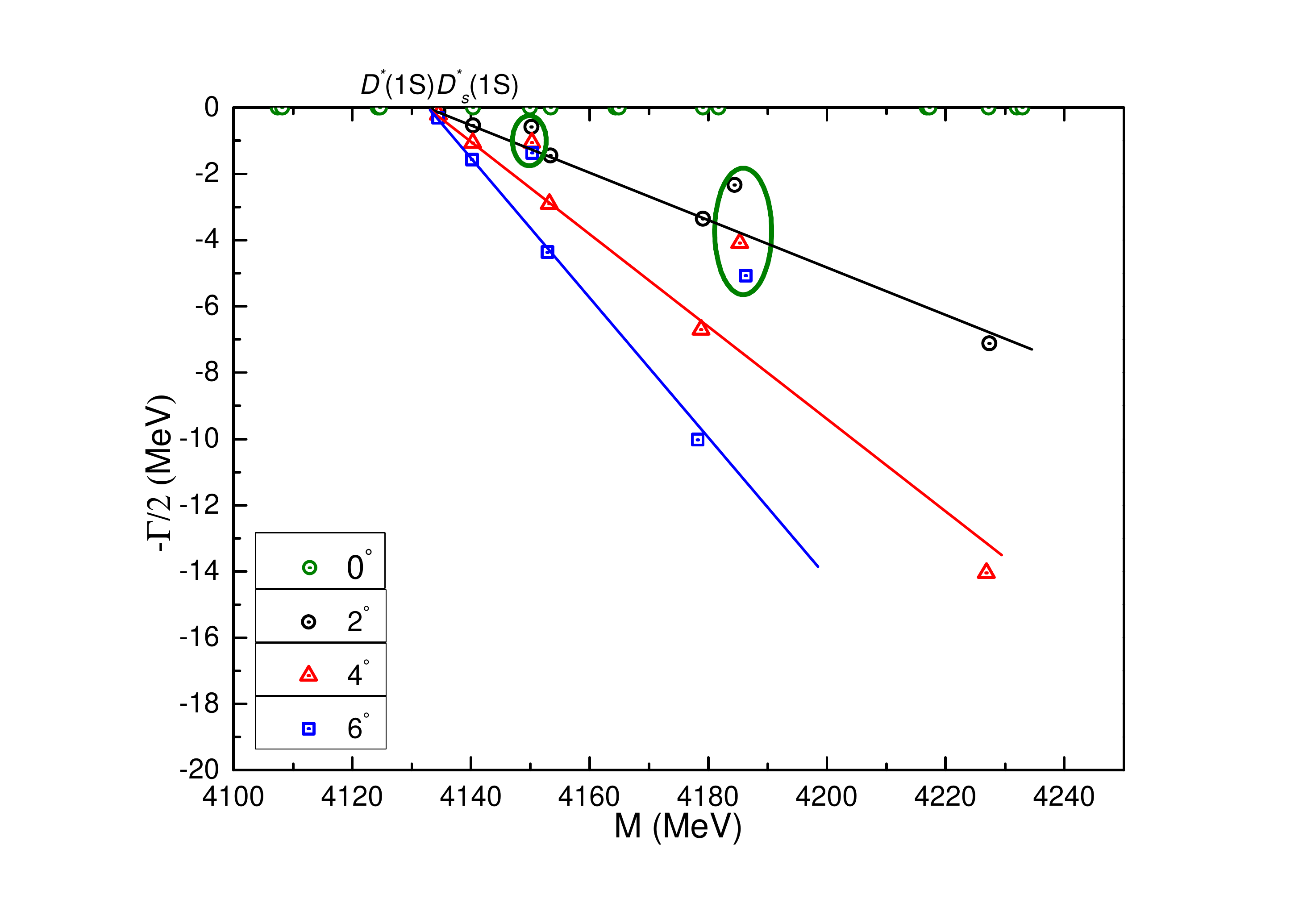} \\
\includegraphics[clip, trim={3.0cm 2.0cm 3.0cm 1.0cm}, width=0.45\textwidth]{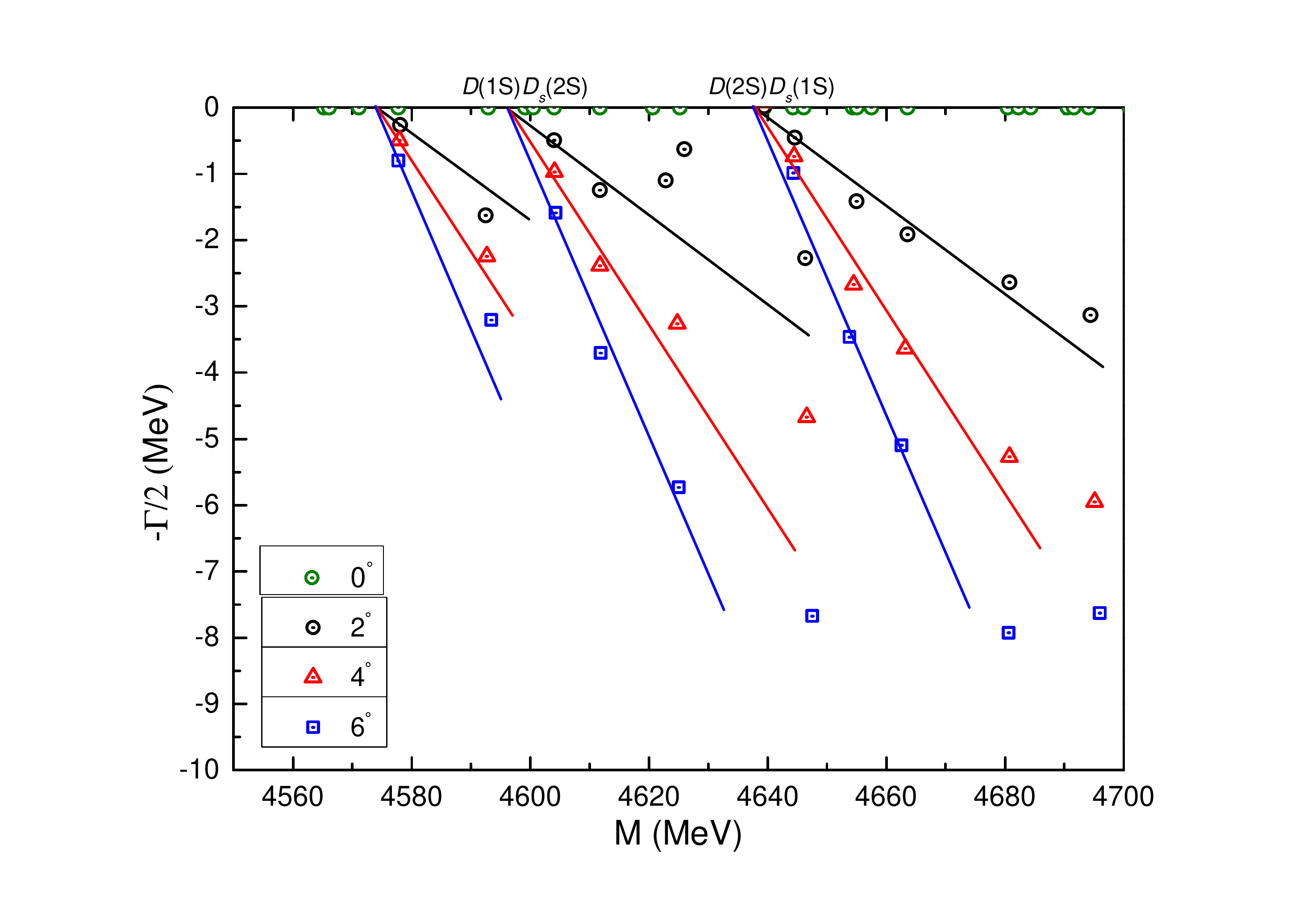}
\caption{\label{PP2} {\it Top panel:} Complex energy spectrum of the $c\bar{c}s\bar{q}$ system with $J^{P}=0^{+}$ from the coupled-channel calculation with CSM. Particularly, the $(c\bar{c})(s\bar{q})$ dimeson channels are excluded. The parameter $\theta$ varies from $0^\circ$ to $6^\circ$. {\it Middle panel:} Enlarged top panel, with real values of energy ranging from $4.10\,\text{GeV}$ to $4.25\,\text{GeV}$. {\it Bottom panel:} Enlarged top panel, with real values of energy ranging from $4.55\,\text{GeV}$ to $4.70\,\text{GeV}$.} 
\end{figure}

\begin{figure}[ht]
\includegraphics[clip, trim={3.0cm 2.0cm 3.0cm 1.0cm}, width=0.45\textwidth]{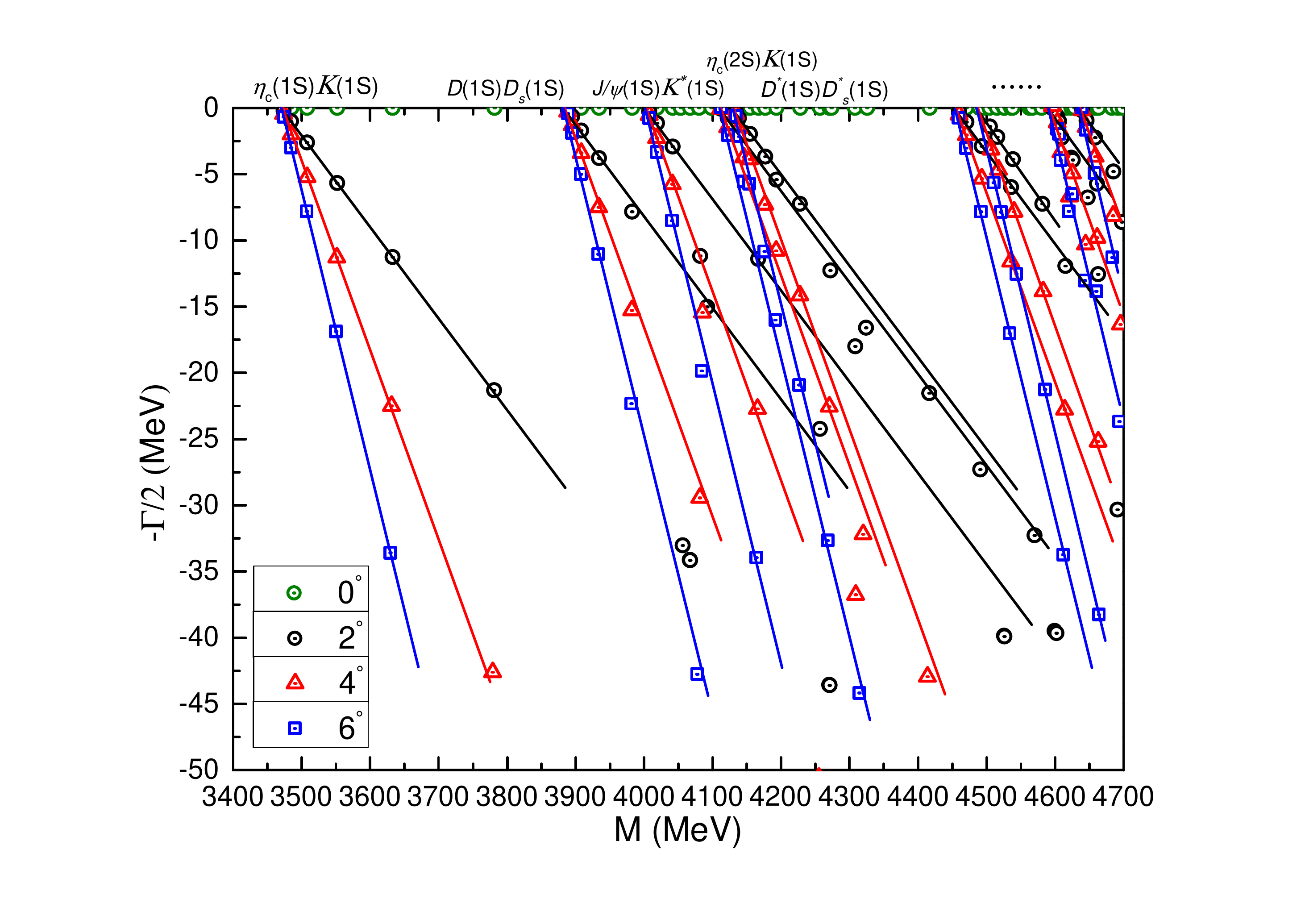} \\
\includegraphics[clip, trim={3.0cm 2.0cm 3.0cm 1.0cm}, width=0.45\textwidth]{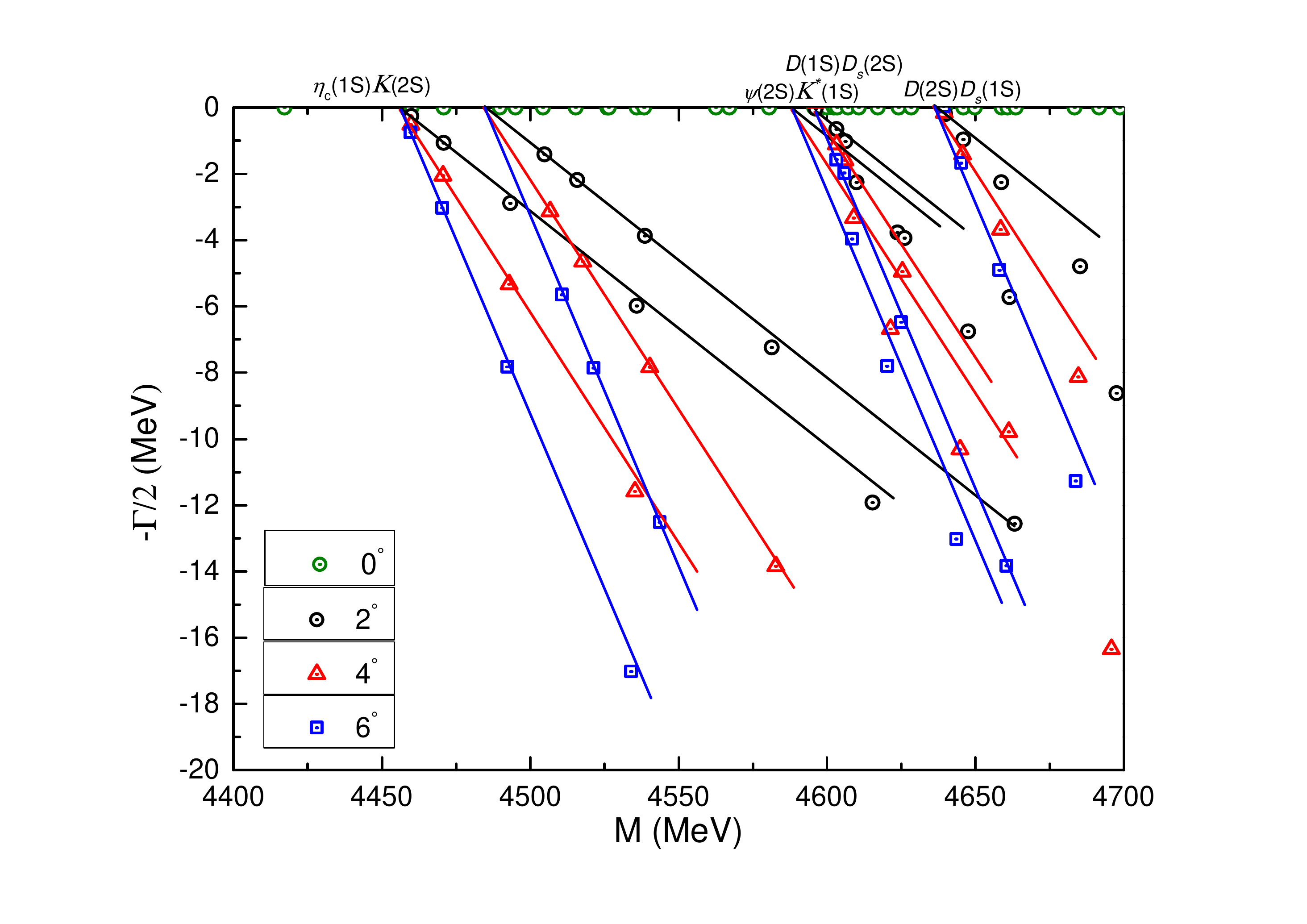} 
\caption{\label{PP3} {\it Top panel:} The complete coupled-channels calculation of the $c\bar{c}s\bar{q}$ tetraquark system with $J^P=0^+$ quantum numbers. We use the complex-scaling method of the chiral quark model varying $\theta$ from $0^\circ$ to $6^\circ$. {\it Bottom panel:} Enlarged top panel, with real values of energy ranging from $4.40\,\text{GeV}$ to $4.70\,\text{GeV}$.}
\end{figure}

Table~\ref{GresultCC1} shows our calculated results of the lowest-lying $c\bar{c}s\bar{q}$ tetraquark states in real-range study. The allowed dimeson, diquark-antidiquark and K-type configurations are listed in the 1st column; when possible, the experimental value of the non-interacting meson-meson threshold is labeled in parentheses. Each channel is assigned an index in the 2nd column. The theoretical mass obtained in each channel is shown in the 3rd column and the coupled-channels result for each kind of configuration is presented in the last one. When a complete coupled-channels calculation is performed, last row of the table indicates the lowest-lying mass. We show in Figs.~\ref{PP1} to~\ref{PP3} the distribution of complex eigen-energies when the CSM is used in the coupled-channels calculation and, therein, the obtained resonance states are indicated inside circles. Furthermore, when all channels listed in Table~\ref{GDD1} are considered excluding the di-meson states in color-singlet configurations, the obtained resonances below 4.3 GeV along with their inner structures are summarized in Table~\ref{tab:dis1}.

{\bf Exotic states in $(c\bar{c})(s\bar{q})$ dimeson channels:} In Table~\ref{GresultCC1} shows that the lowest channel of $c\bar{c}s\bar{q}$ tetraquark with spin-parity $0^+$ is the color-singlet channel of $\eta_c K$ with a theoretical mass of 3470 MeV, which is just the theoretical threshold value and thus bounding is impossible here. This fact also holds for another higher meson-meson channel, $J/\psi K^*$, whose calculated mass is 4004 MeV. Then, the hidden-color structures for $\eta_c K$ and $J/\psi K^*$ have masses 4422 and 4416 MeV, respectively; obviously, these are much deviated from the relevant experimental data.

A coupled-channels within complex-range calculation is performed in a further step. Fig.~\ref{PP1} presents the calculated results, in which the $(c\bar{c})(s\bar{q})$ dimeson, diquark-antidiquark and K-type configurations are considered. In a mass gap from 3.4 to 4.7 GeV, the calculated complex energy dots of $\eta_c(1S)K(1S)$, $J/\psi(1S)K^*(1S)$, $\eta_c(2S)K(1S)$, $\eta_c(1S)K(2S)$ and $\psi(2S)K^*(1S)$ channels are generally well aligned along their corresponding threshold lines. In particular, with a complex angle varied from $0^\circ$ to $6^\circ$, these energy poles are basically moving along the theoretical cut lines, and this fact confirms their nature as scattering states. In Fig.~\ref{PP1} one could find that three resonance poles exist, circled in green, whose masses and widths are $(4255, 5.0)$ MeV, $(4567, 1.6)$ MeV and $(4675, 1.9)$ MeV, respectively. These narrow resonances are obtained in the $(c\bar{c})(s\bar{q})$ dimeson channels along with the couplings in diquark-antidiquark and K-type configurations. Moreover, attending to their positions in the complex plane, the dominant components of them can be identified as $\eta_c(2S)K(1S)(4255)$, $\eta_c(1S)K(2S)(4567)$ and $\psi(2S)K^*(1S)(4675)$, respectively. Accordingly, the reported $Z_{cs}(4220)$ and $X(4630)$ states can be related to the radial excitation of $\eta_c K$ and $\psi K^*$ in coupled-channels cases, without $D^{(*)}D^{(*)}_s$ channel included.

{\bf Exotic states in $(c\bar{q})(s\bar{c})$ dimeson channels:} In a single channel calculation of $(c\bar{q})(s\bar{c})$ meson-meson configuration, the lowest mass 3886 MeV is equal to the theoretical threshold value of $DD_s$ color-singlet channel in Table~\ref{GresultCC1}, and the other $(c\bar{q})(s\bar{c})$ dimeson channel, $D^* D^*_s$, is at 4132 MeV. Threfore, no bound state is found herein too, since the color-octet channels of $DD_s$ and $D^* D^*_s$ are both excited states at around 4.3 GeV.

In a further step, a coupled-channels calculation which includes the $(c\bar{q})(s\bar{c})$ dimeson, diquark-antidiquark and K-type configurations is considered in the CSM. Fig.~\ref{PP2} presents the general distributions of calculated complex energy dots. Particularly, in the top panel of Fig.~\ref{PP2}, most of the complex poles of $D^{(*)}D^{(*)}_s$ are aligned well along the cut lines within a mass region from 3.8 to 4.7 GeV. However, there are dense distributions around 4.2 and 4.6 GeV, and so two enlarged panels are shown accordingly. Firstly, two resonance poles are obtained in the middle panel of Fig.~\ref{PP2} whose mass gap ranges from 4.10 to 4.25 GeV. They can be identified as $D^*(1S)D^*_s(1S)$ narrow resonances with calculated masses and widths $(4150, 2.2)\,MeV$ and $(4185, 8.2)\,MeV$, respectively. For the lower resonance, despite its small decay width, our theoretical mass is compatible with the $Z'_{cs}(4130)$ (tensor $D^*\bar{D}^*_s$ resonance) concluded by Refs.~\cite{Meng:2020ihj, Meng:2021rdg}.

The bottom panel of Fig.~\ref{PP2} shows the highest energy region, 4.55$\sim$4.70 GeV, where the radial excitations of $DD_s$ thresholds are clearly identified. Therein, the calculated complex dots are generally aligned along their corresponding threshold lines, and no stable resonance pole is found.

\begin{table}[!t]
\caption{\label{tab:dis1} The distance, in fm, between any two quarks of the $J^P=0^+$ $c\bar{c}s\bar{q}$ tetraquark resonance state obtained in all exotic configurations' coupled-channels calculation. These resonances, which masses are below 4.3 GeV, are labeled in the first column.}
\begin{ruledtabular}
\begin{tabular}{ccccccc}
  ~~State & $r_{\bar{c}c}$ & $r_{\bar{c}\bar{q}}$ & $r_{\bar{c}s}$ & $r_{c\bar{q}}$  & $r_{cs}$ & $r_{s\bar{q}}$~~  \\[2ex]
  ~~$Z_{cs}(3841)$& 0.35 & 0.76 & 0.70 & 0.76 & 0.70 & 0.69 ~~ \\
  ~~$Z_{cs}(4105)$& 0.51 & 0.84 & 0.69 & 0.77 & 0.76 & 0.85 ~~ \\
  ~~$Z_{cs}(4156)$& 0.42 & 0.98 & 0.92 & 0.98 & 0.91 & 0.89 ~~ \\
  ~~$Z_{cs}(4193)$& 0.42 & 1.16 & 1.13 & 1.15 & 1.14 & 1.02 ~~ \\
  ~~$Z_{cs}(4258)$& 0.64 & 0.82 & 0.63 & 0.73 & 0.73 & 0.91 ~~ \\
\end{tabular}
\end{ruledtabular}
\end{table}

{\bf The fully coupled-channels case:} There are 28 channels under consideration for the $c\bar{c}s\bar{q}$ tetraquark with spin-parity $0^+$. In Table~\ref{GresultCC1} one could see that, apart from the dimeson channels in both color-singlet and -octet cases, four allowed channels in diquark-antidiquark and K-type configurations are also considered. Generally, in each single channel computation, masses of diquark-antidiquark structures are about 4.3 GeV, and 4.1$\sim$4.4 GeV for K-type ones. Then, in coupled-channels calculations of each specific configurations, the lowest mass, 3470 MeV, is still just the theoretical threshold value of $\eta_c K$, and the other exotic structures are around 4.1 GeV. In particular, the coupled masses in $K_1$ and $K_2$ structures are both $\sim$4.05 GeV, which is close to the experimental mass of $Z_{cs}(4000)$. Finally, within the real-range investigation, the fully coupled-channels mass remains at 3470 MeV which implies the coupling effect is quite weak and no bound state is available.

The exotic state within hidden-color channel could be a natural bound state, hence a coupled-channels calculation of all those exotic structures which include color-octet, diquark-antidiquark and K-type configurations are performed. Table~\ref{tab:dis1} lists the obtained $c\bar{c}s\bar{q}$ tetraquark resonances below 4.3 GeV, and their inner structures are also investigated. The lowest resonance is at 3841 MeV, and the other four excited states are lying in 4.11$\sim$4.25 GeV. Apparently, these exotic resonance states are likely compact structures with inter-quark distance generally less than 1.1 fm, and the distance between $c\bar{c}$ is around 0.5 fm. They could be good candidates of exotic color structures in the strange hidden-charm tetraquark sector.

Additionally, the fully coupled-channels calculation is studied in a complex-range scaling. Fig.~\ref{PP3} shows the distribution of our calculated complex energies. Particularly, in the top panel, the scattering states of $\eta_c K$, $J/\psi K^*$, $DD_s$ and $D^* D^*_s$ can be clearly identified within the mass region 3.4$\sim$4.7 GeV. Besides, since there is a dense distribution between 4.5 and 4.7 GeV, an enlarged part whose mass region is from 4.4 to 4.7 GeV is shown in the bottom panel of Fig.~\ref{PP3}. Therein, the scattering nature of the radial excitated states $\eta_c K$, $\psi K^*$ and $DD_s$ is also well presented. One can realize that the obtained resonance states in the coupled-channels investigations discussed above, $\eta_c(2S)K(1S)(4255)$, $\psi(2S)K^*(1S)(4675)$, $D^*(1S)D^*_s(1S)(4150)$, etc., turn to be scattering ones because, in a complete coupled-channels calculation, these resonances easily decay to lower $(c\bar{c})(s\bar{q})$ or $(c\bar{q})(s\bar{c})$ di-meson scattering states.


\begin{table*}[!t]
\caption{\label{GresultCC2} Lowest-lying $c\bar{c}s\bar{q}$ tetraquark states with $J^P=1^+$ calculated within the real range formulation of the chiral quark model.
The allowed meson-meson, diquark-antidiquark and K-type configurations are listed in the first and fifth columns; when possible, the experimental value of the non-interacting meson-meson threshold is labeled in parentheses. Each channel is assigned an index in the 2nd and 6th columns. The theoretical mass obtained in each channel is shown in the 3rd and 7th columns and the coupled result for each kind of configuration is presented in the 4th and last columns.
When a complete coupled-channels calculation is performed, last row of the table indicates the lowest-lying mass.
(unit: MeV).}
\begin{ruledtabular}
\begin{tabular}{cccccccc}
Channel~ & Index & $M$ & Mixed~~ & Channel~~ & Index & $M$  & Mixed~~ \\[2ex]
 $(\eta_c K^*)^1(3873)$ & 1 & $3896$ &   & $K_1$ & 19 & $4426$ & \\
 $(J/\psi K)^1(3591)$ & 2 & $3578$  &  & $K_1$ & 20 & $4422$ & \\
 $(J/\psi K^*)^1(3989)$ & 3  & $4004$ &   & $K_1$ & 21 & $4443$ & \\
 $(DD^*_s)^1(3982)$ & 4 & $4012$  &  & $K_1$ & 22 & $4231$ &   \\
 $(D^* D_s)^1(3975)$  & 5 & $4006$  &  & $K_1$ & 23 & $4266$ & \\
 $(D^* D^*_s)^1(4119)$ & 6  & $4132$  & $3578$  & $K_1$  & 24 & $4188$ & $4186$ \\[2ex]
 $(\eta_c K^*)^8$ & 7 & $4448$ &   & $K_2$  & 25 & $4215$ & \\
 $(J/\psi K)^8$ & 8 & $4424$ &  & $K_2$  & 26 & $4252$ & \\
 $(J/\psi K^*)^8$ & 9  & $4434$ &  & $K_2$  & 27 & $4175$ & \\
 $(DD^*_s)^8$ & 10 & $4403$ &  & $K_2$  & 28 & $4420$ &  \\
 $(D^* D_s)^8$  & 11 & $4401$  &  & $K_2$  & 29 & $4427$ & \\
 $(D^* D^*_s)^8$ & 12  & $4361$ & $4224$  & $K_2$  & 30 & $4442$ & $4168$ \\[2ex]
 $(cs)_3 (\bar{c}\bar{q})^*_{\bar{3}}$ & 13 & $4352$ & & $K_3$  & 31 & $4283$ & \\
 $(cs)^*_3 (\bar{c}\bar{q})_{\bar{3}}$  & 14 & $4357$ & & $K_3$  & 32 & $4306$ & \\
 $(cs)^*_3 (\bar{c}\bar{q})^*_{\bar{3}}$ & 15 & $4350$ & & $K_3$  & 33 & $4340$ &\\
 $(cs)_6 (\bar{c}\bar{q})^*_{\bar{6}}$  & 16 & $4342$  & & $K_3$  & 34 & $4314$ &  \\
 $(cs)^*_6 (\bar{c}\bar{q})_{\bar{6}}$   & 17 & $4339$ &  & $K_3$  & 35 & $4348$ & \\
 $(cs)^*_6 (\bar{c}\bar{q})^*_{\bar{6}}$  & 18  & $4270$ & $4184$  & $K_3$  & 36 & $4349$ & $4178$ \\[2ex]
  &  &  &  & $K_4$  & 37  & $4309$   & \\
  &  &  &  & $K_4$  & 38  & $4319$   & \\
  &  &  &  & $K_4$  & 39  & $4362$   & \\
  &  &  &  & $K_4$  & 40  & $4327$   & \\
  &  &  &  & $K_4$  & 41  & $4353$    & \\
  &  &  &  & $K_4$  & 42  & $4349$   & $4191$ \\[2ex]
\multicolumn{7}{c}{Complete coupled-channels:} & $3578$ \\
\end{tabular}
\end{ruledtabular}
\end{table*}

\subsection{The $\mathbf{J^P=1^+}$ $\mathbf{c\bar{c}s\bar{q}}$ tetraquark system}

\begin{figure}[ht]
\includegraphics[clip, trim={3.0cm 2.0cm 3.0cm 1.0cm}, width=0.45\textwidth]{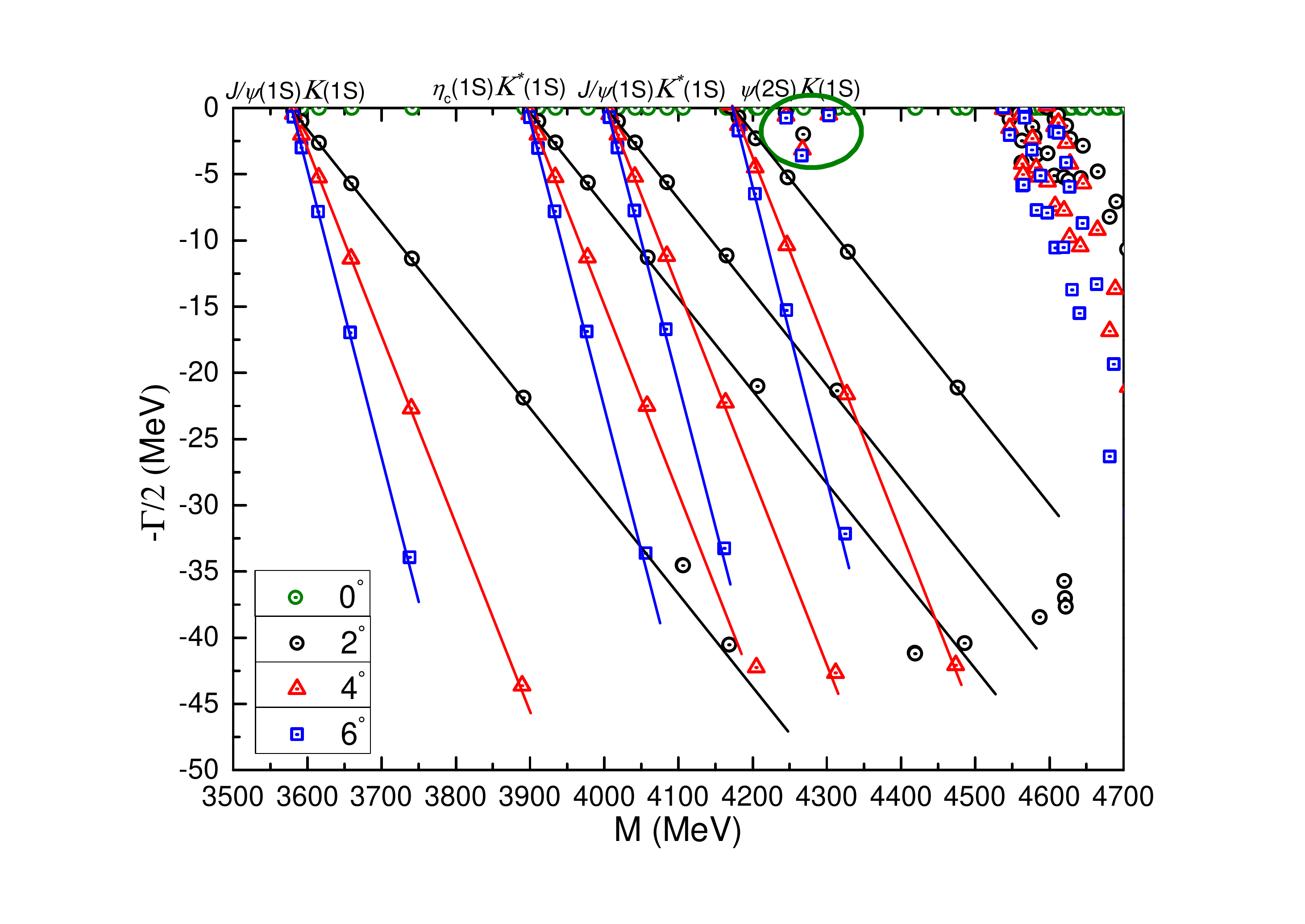} \\
\includegraphics[clip, trim={3.0cm 2.0cm 3.0cm 1.0cm}, width=0.45\textwidth]{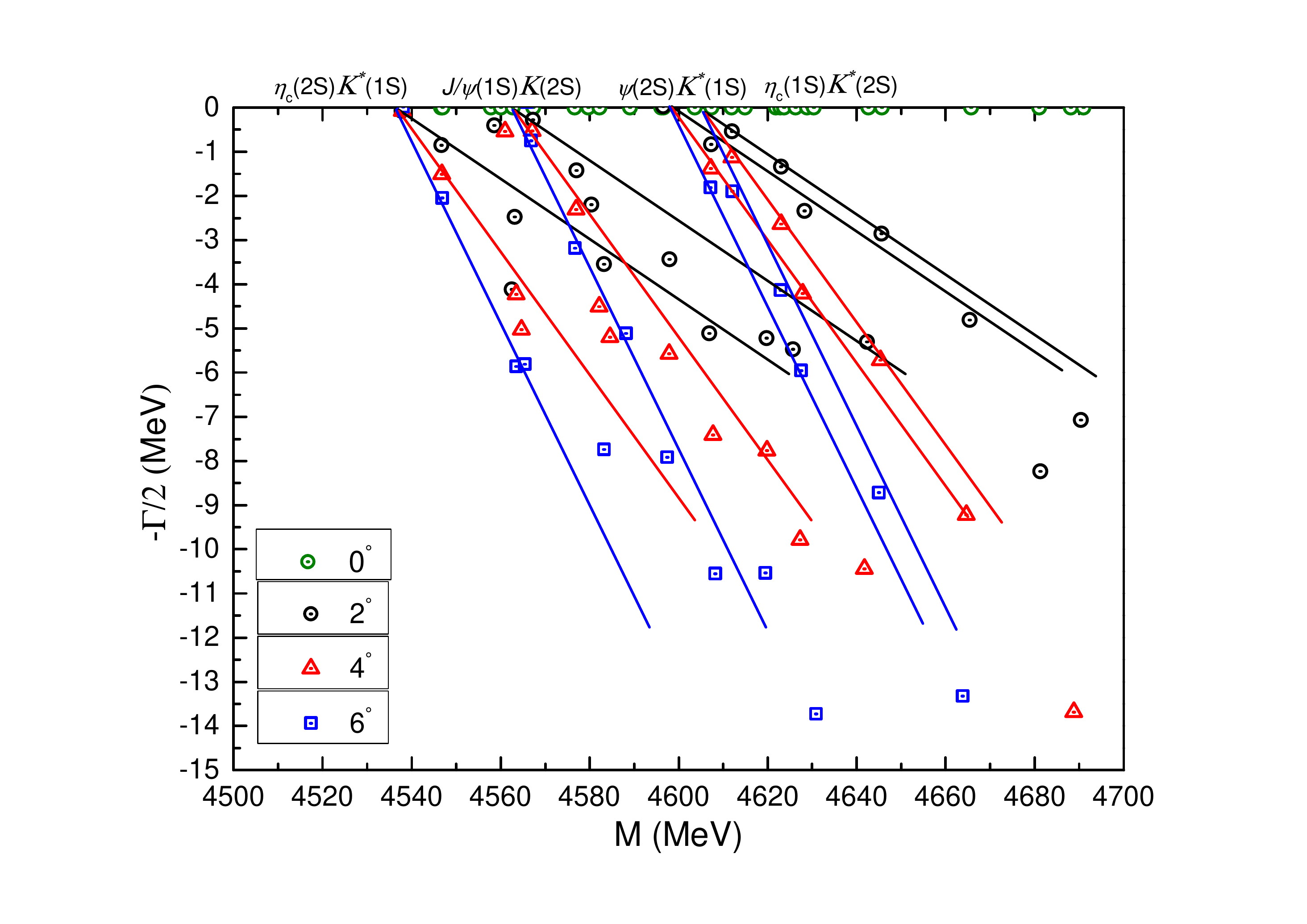}
\caption{\label{PP4} {\it Top panel:} The coupled-channels calculation of the $c\bar{c}s\bar{q}$ tetraquark system with $J^P=1^+$ quantum numbers. Particularly, the $(c\bar{q})(s\bar{c})$ dimeson channels are excluded. We use the complex-scaling method of the chiral quark model varying $\theta$ from $0^\circ$ to $6^\circ$. {\it Bottom panel:} Enlarged top panel, with real values of energy ranging from $4.50\,\text{GeV}$ to $4.70\,\text{GeV}$.}
\end{figure}

\begin{figure}[ht]
\includegraphics[width=0.49\textwidth, trim={2.3cm 2.0cm 2.0cm 1.0cm}]{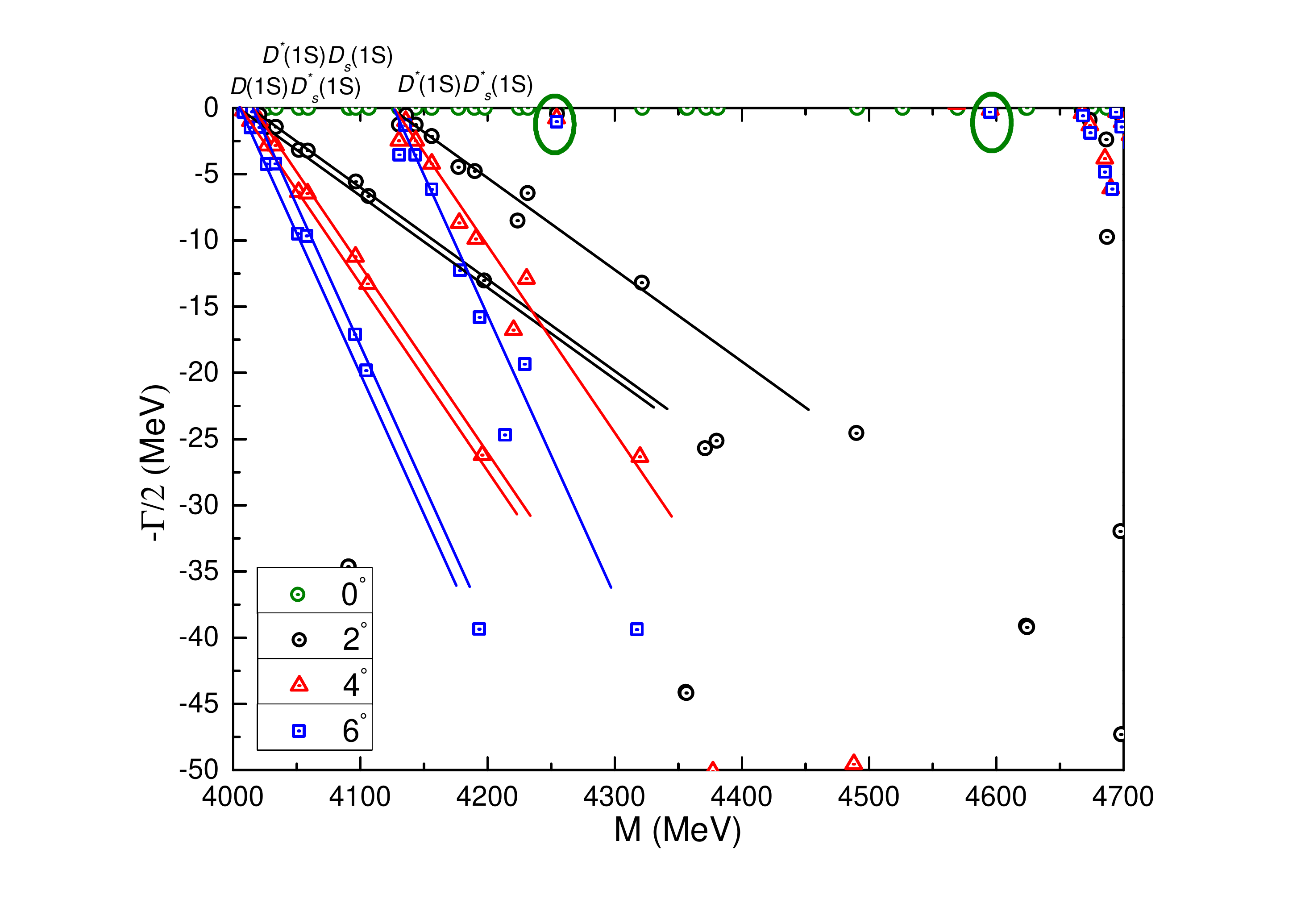}
\caption{\label{PP5} The coupled-channels calculation of the $c\bar{c}s\bar{q}$ tetraquark system with $J^P=1^+$ quantum numbers. Particularly, the $(c\bar{c})(s\bar{q})$ dimeson channels are excluded. We use the complex-scaling method of the chiral quark model varying $\theta$ from $0^\circ$ to $6^\circ$.}
\end{figure}

\begin{figure}[ht]
\includegraphics[clip, trim={3.0cm 2.0cm 3.0cm 1.0cm}, width=0.45\textwidth]{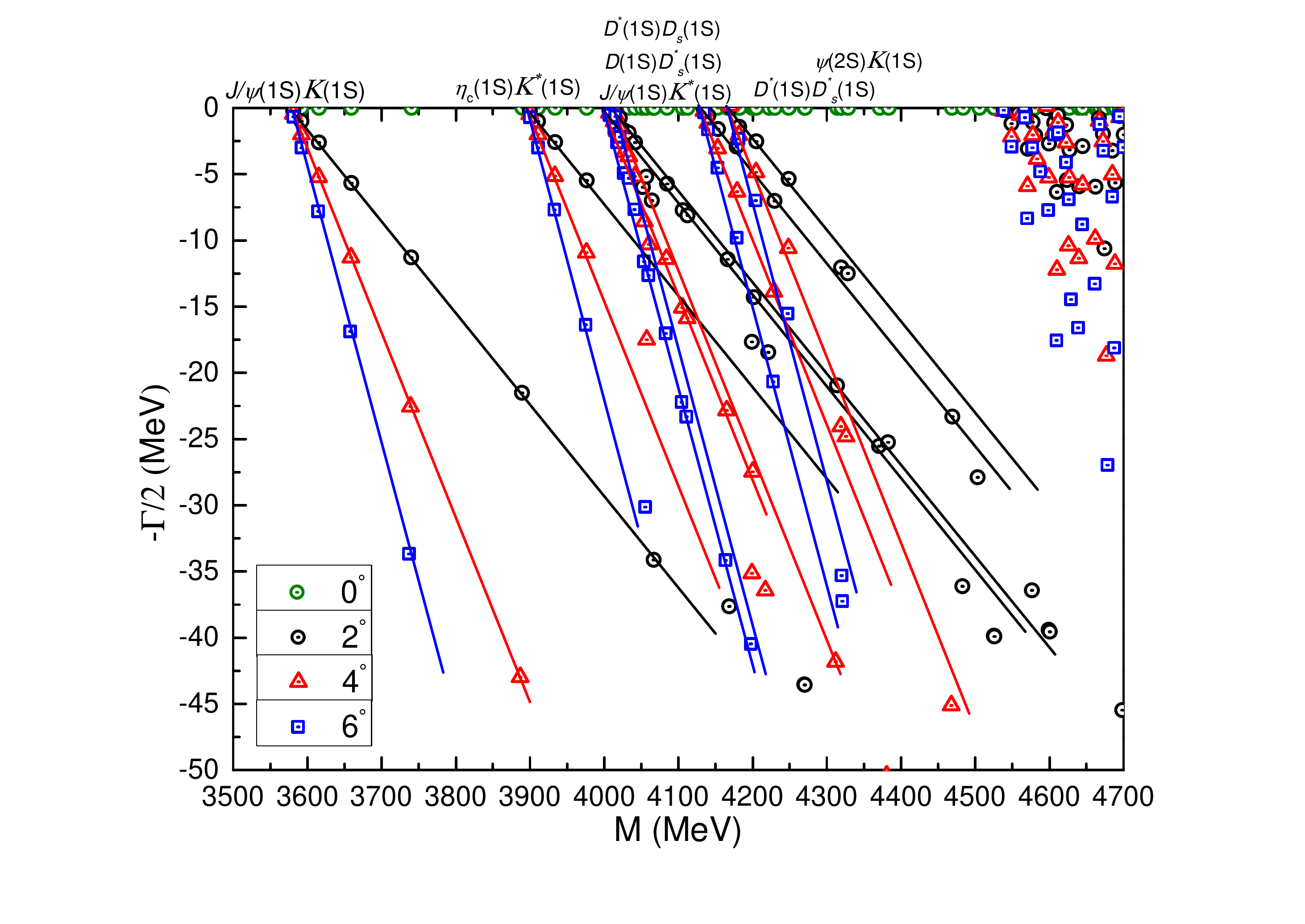} \\
\includegraphics[clip, trim={3.0cm 2.0cm 3.0cm 1.0cm}, width=0.45\textwidth]{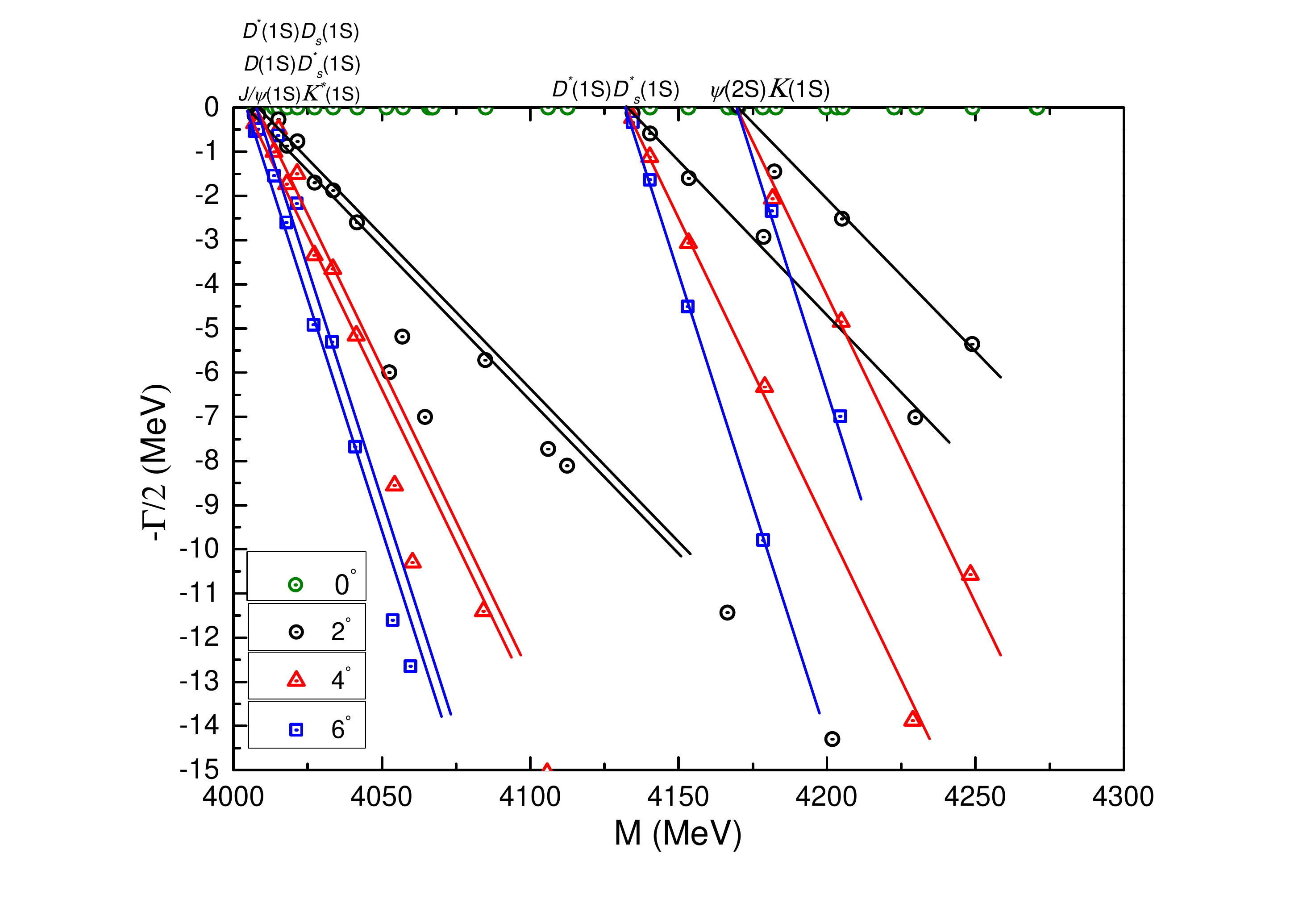} \\
\includegraphics[clip, trim={3.0cm 2.0cm 3.0cm 1.0cm}, width=0.45\textwidth]{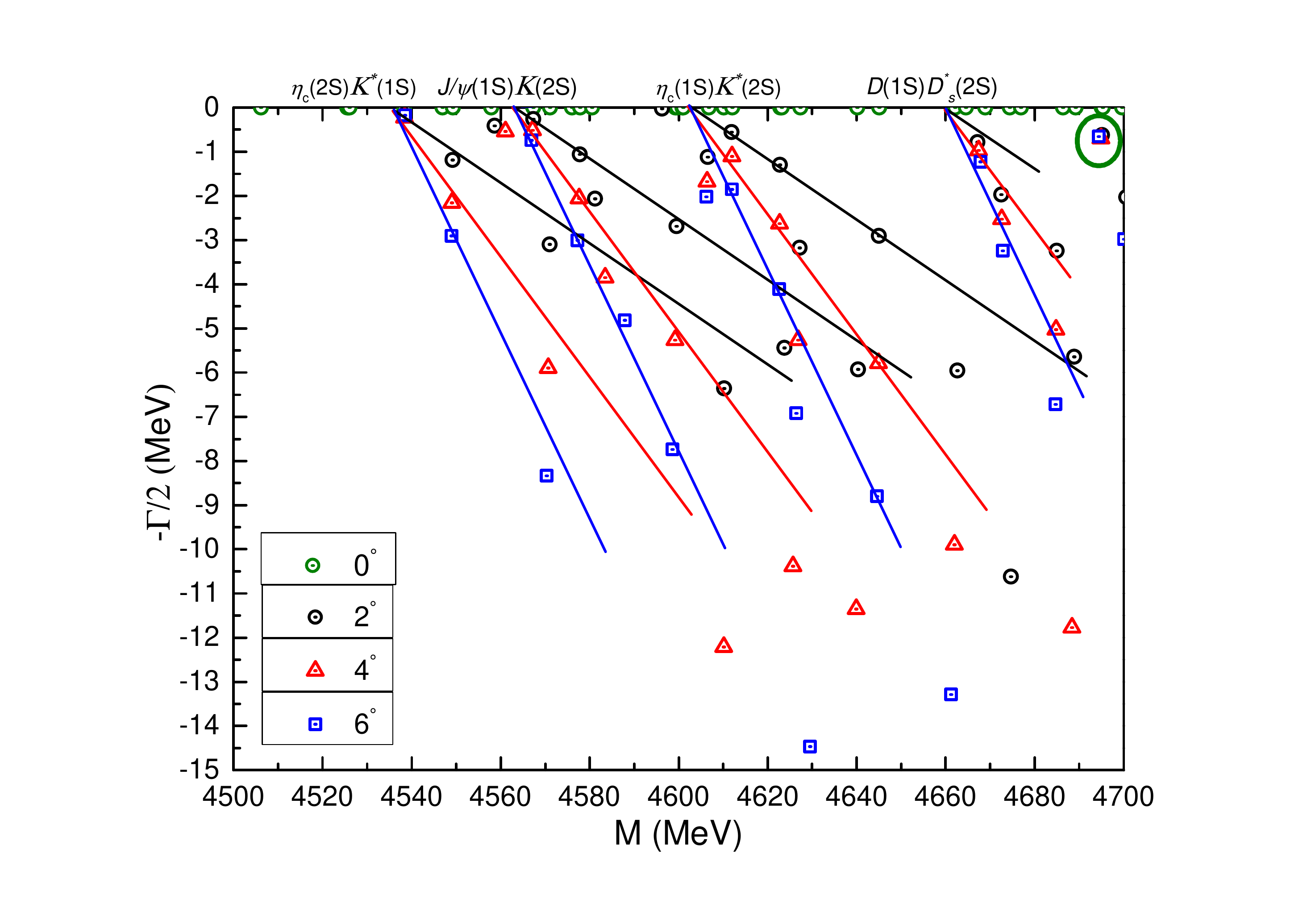}
\caption{\label{PP6} {\it Top panel:} The complete coupled-channels calculation of the $c\bar{c}s\bar{q}$ tetraquark system with $J^P=1^+$ quantum numbers. We use the complex-scaling method of the chiral quark model varying $\theta$ from $0^\circ$ to $6^\circ$. {\it Middle panel:} Enlarged top panel, with real values of energy ranging from $4.00\,\text{GeV}$ to $4.30\,\text{GeV}$. {\it Bottom panel:} Enlarged top panel, with real values of energy ranging from $4.50\,\text{GeV}$ to $4.70\,\text{GeV}$.}
\end{figure}

The available 42 channels that include meson-meson, diquark-antidiquark and K-type structures are listed in Table~\ref{GDD2}. The lowest-lying $c\bar{c}s\bar{q}$ tetraquark states with spin-parity $J^P=1^+$ are firslty calculated within the real-range approximation and summarized in Table~\ref{GresultCC2}. Particularly, the allowed channels are listed in the 1st and 5th columns, when possible, the non-interacting meson-meson experimental threshold values are labeled in parentheses. The assigned indexes for the channels are shown in the 2nd and 6th columns. The theoretical mass obtained in each channel is shown in the 3rd and 7th columns, besides, the coupled result for each kind of configuration is presented in the 4th and last columns. The last row of Table~\ref{GresultCC2} indicates the lowest-lying mass of the system in a fully coupled-channels case. When the CSM is used in the coupled-channels calculations, Figs.~\ref{PP4} to~\ref{PP6} show the distribution of complex eigen-energies and, therein, the obtained resonance states are indicated inside circles. Furthermore, when a calculation which all channels listed in Table~\ref{GDD2} are considered except the color-singlet di-meson cases is performed, the obtained resonances within a mass region 3.9$\sim$4.2 GeV, along with their inner structures, are summarized in Table~\ref{tab:dis2}. Now let us discuss the details of $c\bar{c}s\bar{q}$ tetraquark in each three kind of investigations below.

{\bf Exotic states in $(c\bar{c})(s\bar{q})$ dimeson channels:} There are six $(c\bar{c})(s\bar{q})$ di-meson channels contributing to the tetraquark system in $1^+$ case, \emph{i.e.}, $\eta_c K^*$, $J/\psi K$ and $J/\psi K^*$ states in color-singlet and -octet channels. From Table~\ref{GresultCC2}, we find that the three color-singlet channels are all unbound, the lowest-lying one, $J/\psi K$, is at 3578 MeV, the next one is $\eta_c K^*$ with mass at 3896 MeV, and $J/\psi K^*$ is at 4004 MeV. Their corresponding hidden-color channels masses are all $\sim$4.4 GeV.

Additionally, a complex-range computation is performed in the coupled-channels study where the $(c\bar{c})(s\bar{q})$ dimeson structures, diquark-antidiquark configurations and K-type ones are considered. Within a complex angle $\theta$ varied from $0^\circ$ to $6^\circ$, Fig.~\ref{PP4} shows the distribution of calculated complex energy dots. In particular, the scattering states of $J/\psi K$, $\eta_c K^*$ and $J/\psi K^*$ are well presented in a mass region 3.5$\sim$4.7 GeV of the top panel of Fig.~\ref{PP4}. However, there are three stable resonance poles, encircled by green lines, which can be identified as $\psi(2S)K(1S)$ molecular resonances. Their calculated masses and widths are, all in MeV, $(4254, 0.8)$, $(4267, 6.2)$ and $(4303, 1.0)$. As one could conclude, the $Z_{cs}(4220)$ would be explained, too, as a $\psi(2S)K(1S)(4254)$ resonance in $1^+$ state.

Meanwhile, since there is a dense distribution of complex energies at around 4.6 GeV, an enlarged panel whose mass range goes from 4.5 to 4.7 GeV is shown at the bottom of Fig.~\ref{PP4}. Therein, the radial excitation states of $\eta_c K^*$ and $J/\psi K^{(*)}$ are clearly presented and no resonance pole is obtained.

{\bf Exotic states in $(c\bar{q})(s\bar{c})$ dimeson channels:} One can find in Table~\ref{GresultCC2} that there are six $(c\bar{q})(s\bar{c})$ dimeson channels which contribute to the tetraquark system with $1^+$ quantum numbers, \emph{i.e.}, $DD^*_s$, $D^* D_s$ and $D^* D^*_s$ states in color-singlet and -octet channels. Within the real-range single channel approach, their theoretical masses in color-singlet channels are $\sim$4.0 GeV, except for $D^* D^*_s$ at 4132 MeV. The hidden-color channels masses are higher and generally lie at around 4.4 GeV. Although these values are quite close to the experimental data of $Z_{cs}(4000)$, the scattering nature of $D^{(*)}D^{(*)}_s$ channel remains.

The coupled-channels result by CSM is shown in Fig.~\ref{PP5}, within a mass range from 4.0 to 4.7 GeV. The scattering nature of $DD^*_s$ and $D^* D_s$ is clearly demonstrated. Nevertheless, two narrow $D^*(1S) D^*_s(1S)$ resonances are obtained and circled in green. The calculated masses and widths are $(4254, 1.6)$ MeV and $(4594, 0.6)$ MeV, respectively. Accordingly, there is a degeneration between the $D^*(1S) D^*_s(1S)$ and $\psi(2S)K(1S)$ channels at 4254 MeV.

\begin{table}[!t]
\caption{\label{tab:dis2} The distance, in fm, between any two quarks of the $J^P=1^+$ $c\bar{c}s\bar{q}$ tetraquark resonance state obtained in all exotic configurations' coupled-channels calculation. These resonances, which masses are around 3.9$\sim$4.2 GeV, are labeled in the first column.}
\begin{ruledtabular}
\begin{tabular}{ccccccc}
  ~~State & $r_{\bar{c}c}$ & $r_{\bar{c}\bar{q}}$ & $r_{\bar{c}s}$ & $r_{c\bar{q}}$  & $r_{cs}$ & $r_{s\bar{q}}$~~  \\[2ex]
  ~~$Z_{cs}(3947)$& 0.41 & 0.78 & 0.71 & 0.77 & 0.71 & 0.70 ~~ \\
  ~~$Z_{cs}(4038)$& 0.35 & 0.95 & 0.87 & 0.95 & 0.87 & 0.87 ~~ \\
  ~~$Z_{cs}(4137)$& 0.44 & 0.94 & 0.83 & 0.92 & 0.86 & 0.88 ~~ \\
  ~~$Z_{cs}(4199)$& 0.58 & 0.80 & 0.67 & 0.74 & 0.73 & 0.87 ~~ \\
  ~~$Z_{cs}(4211)$& 0.60 & 0.84 & 0.65 & 0.76 & 0.75 & 0.90 ~~ \\
\end{tabular}
\end{ruledtabular}
\end{table}

{\bf The fully coupled-channels case:} Apart from the 12 meson-meson channels listed in Table~\ref{GresultCC2}, the single channel calculation for diquark-antidiquark structures and K-type ones is also analyzed. In particular, there are six channels in each kind of configurations, the diquark-antidiquark channels are $\sim$4.3 GeV, and the K-type ones are lying from 4.2 to 4.4 GeV. In addition, when a coupled-channels study is done in each configuration, the lowest mass, 3578 MeV, is still equal to the theoretical threshold value of $J/\psi K$. The mass of the hidden-color coupled-channels result is higher and at 4224 MeV. Meanwhile, the diquark-antidiquark and K-type structures coupled masses are all around 4.18 GeV. However, no bound state is found in $1^+$ state when a complete coupled-channels calculation is studied and the lowest mass remains at 3578 MeV.

In order to have a much more detailed investigation on the strange hidden-charm tetraquark in $1^+$ state, a coupled-channels computation of all exotic structures where only the dimeson structures in color-singlet channels are excluded is performed. Table~\ref{tab:dis2} summarizes the $Z_{cs}$ resonances within 3.9$\sim$4.2 GeV. The five exotic states are compact structures and the calculated distances between any two (anti)quark and quark-antiquark are less than 1.0 fm. Particularly, the two lower states, $Z_{cs}(3947)$ and $Z_{cs}(4038)$, are well compatible with the recently reported exotic structures~\cite{BESIII:2020qkh, LHCb:2021uow}. Hence, the $Z_{cs}(3985)$ and $Z_{cs}(4000)$ should be considered as $J^P=1^+$ $c\bar{c}s\bar{q}$ tetraquark with a large compact component.

Last but not least, a fully coupled-channels calculation in CSM is investigated and the result is listed in Fig.~\ref{PP6}. Firstly, within a mass region 3.5$\sim$4.7 GeV, the scattering states of $J/\psi K^{(*)}$, $\eta_c K^*$ and $D^{(*)}D^*_s$ are clearly shown in the top panel of Fig.~\ref{PP6}. One could see that with the rotated angle $\theta$ varied from $0^\circ$ to $6^\circ$, the calculated complex energy dots always descend and basically align along the corresponding threshold lines. However, since there are dense distributions at around 4.1 and 4.6 GeV, enlarged parts on these two energy regions are presented in, respectively, the middle and bottom panels of Fig.~\ref{PP6}. No resonance pole is obtained in the energy range 4.0$\sim$4.3 GeV of the middle panel; however, one narrow resonance is found at higher energy, shown in the bottom panel. With a mass range from 4.5 to 4.7 GeV, four continuum (scattering) states of $\eta_c K^*$, $J/\psi K$ and $DD^*_s$ in radial excitations are well shown; meanwhile, a fixed pole is circled at $(4695, 1.3)$ MeV. Accordingly, the $X(4685)$ state with $1^+$ quantum numbers reported by the LHCb collaboration~\cite{LHCb:2021uow} could be explained as a $D(1S)D^*_s(2S)(4695)$ resonance.

It is worth emphasizing herein that the resonances obtained in the different kinds of coupled-channels investigations are quite unstable and they easily decay to a meson-meson scattering state, \emph{e.g.}, the $Z_{cs}(3947)$, $Z_{cs}(4038)$ and $Z_{cs}(4137)$, which are obtained in a coupled-channels calculation with only exotic color structures are unavailable in the complete coupled-channels case of Fig.~\ref{PP6}.


\begin{table}[!t]
\caption{\label{GresultCC3} Lowest-lying $c\bar{c}s\bar{q}$ tetraquark states with $J^P=2^+$ calculated within the real range formulation of the chiral quark model.
The allowed meson-meson, diquark-antidiquark and K-type configurations are listed in the first column; when possible, the experimental value of the non-interacting meson-meson threshold is labeled in parentheses. Each channel is assigned an index in the 2nd column. The theoretical mass obatined in each channel is shown in the 3rd column and the coupled result for each kind of configuration is presented in the last column.
When a complete coupled-channels calculation is performed, last row of the table indicates the lowest-lying mass.
(unit: MeV).}
\begin{ruledtabular}
\begin{tabular}{lccc}
~~Channel   & Index & $M$ & Mixed~~ \\[2ex]
$(J/\psi K^*)^1 (3989)$  & 1    & $4004$ & \\
$(D^* D^*_s)^1 (4119)$  & 2     & $4132$ & $4004$ \\[2ex]
$(J/\psi K^*)^8$  & 3     & $4466$ &  \\
$(D^* D^*_s)^8$  & 4     & $4411$ & $4317$ \\[2ex]
$(cs)^*_3 (\bar{c}\bar{q})^*_{\bar{3}}$      & 5    & $4381$ & \\
$(cs)^*_6 (\bar{c}\bar{q})^*_{\bar{6}}$  & 6     & $4346$ & $4296$ \\[2ex]
$K_1$  & 7     & $4462$ & \\
  & 8     & $4296$ & $4296$ \\[2ex]
$K_2$  & 9     & $4283$ & \\
  & 10     & $4461$ & $4283$ \\[2ex]
$K_3$  & 11    & $4348$ & \\
  & 12     & $4373$ & $4302$ \\[2ex]
$K_4$  & 13     & $4367$ & \\
      & 14     & $4378$ & $4302$ \\[2ex]
\multicolumn{3}{c}{Complete coupled-channels:} & $4004$ \\
\end{tabular}
\end{ruledtabular}
\end{table}

\subsection{The $\mathbf{J^P=2^+}$ $\mathbf{c\bar{c}s\bar{q}}$ tetraquark system}

\begin{figure}[ht]
\includegraphics[width=0.49\textwidth, trim={2.3cm 2.0cm 2.0cm 1.0cm}]{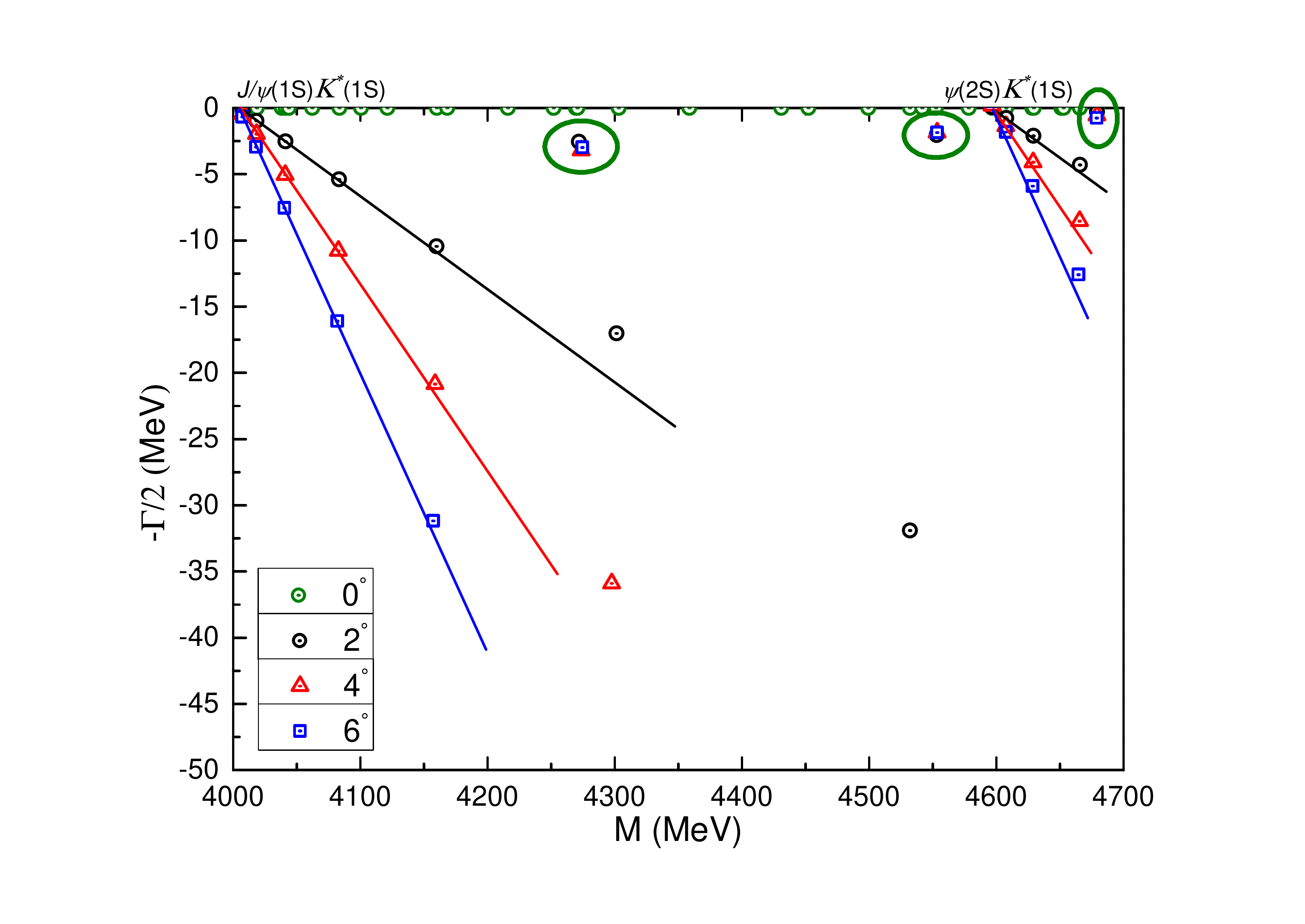}
\caption{\label{PP7} The coupled-channels calculation of the $c\bar{c}s\bar{q}$ tetraquark system with $J^P=2^+$ quantum numbers. Particularly, the $(c\bar{q})(s\bar{c})$ dimeson channels are excluded. We use the complex-scaling method of the chiral quark model varying $\theta$ from $0^\circ$ to $6^\circ$.}
\end{figure}

\begin{figure}[ht]
\includegraphics[width=0.49\textwidth, trim={2.3cm 2.0cm 2.0cm 1.0cm}]{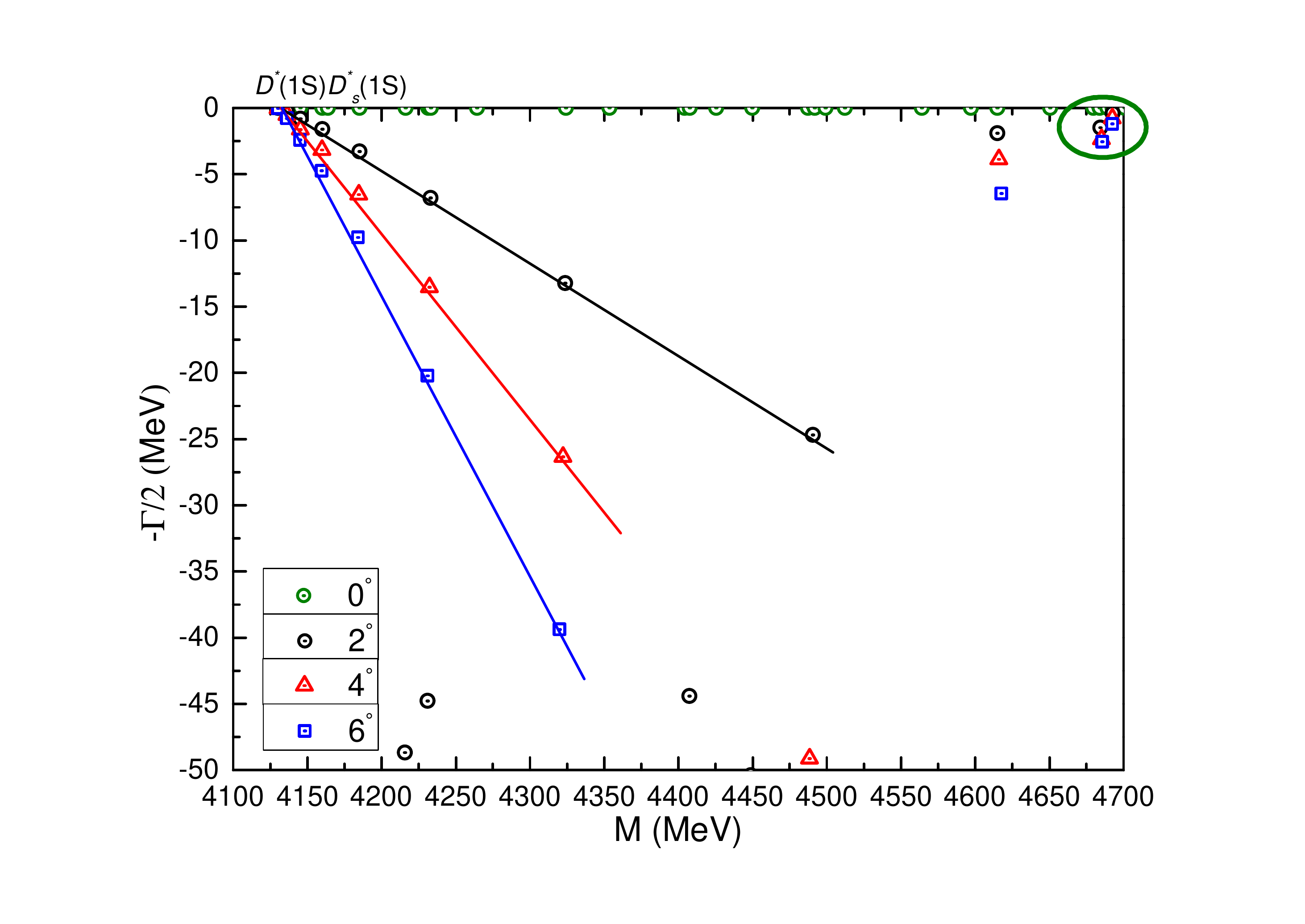}
\caption{\label{PP8} The coupled-channels calculation of the $c\bar{c}s\bar{q}$ tetraquark system with $J^P=2^+$ quantum numbers. Particularly, the $(c\bar{c})(s\bar{q})$ dimeson channels are excluded. We use the complex-scaling method of the chiral quark model varying $\theta$ from $0^\circ$ to $6^\circ$.}
\end{figure}

\begin{figure}[ht]
\includegraphics[width=0.49\textwidth, trim={2.3cm 2.0cm 2.0cm 1.0cm}]{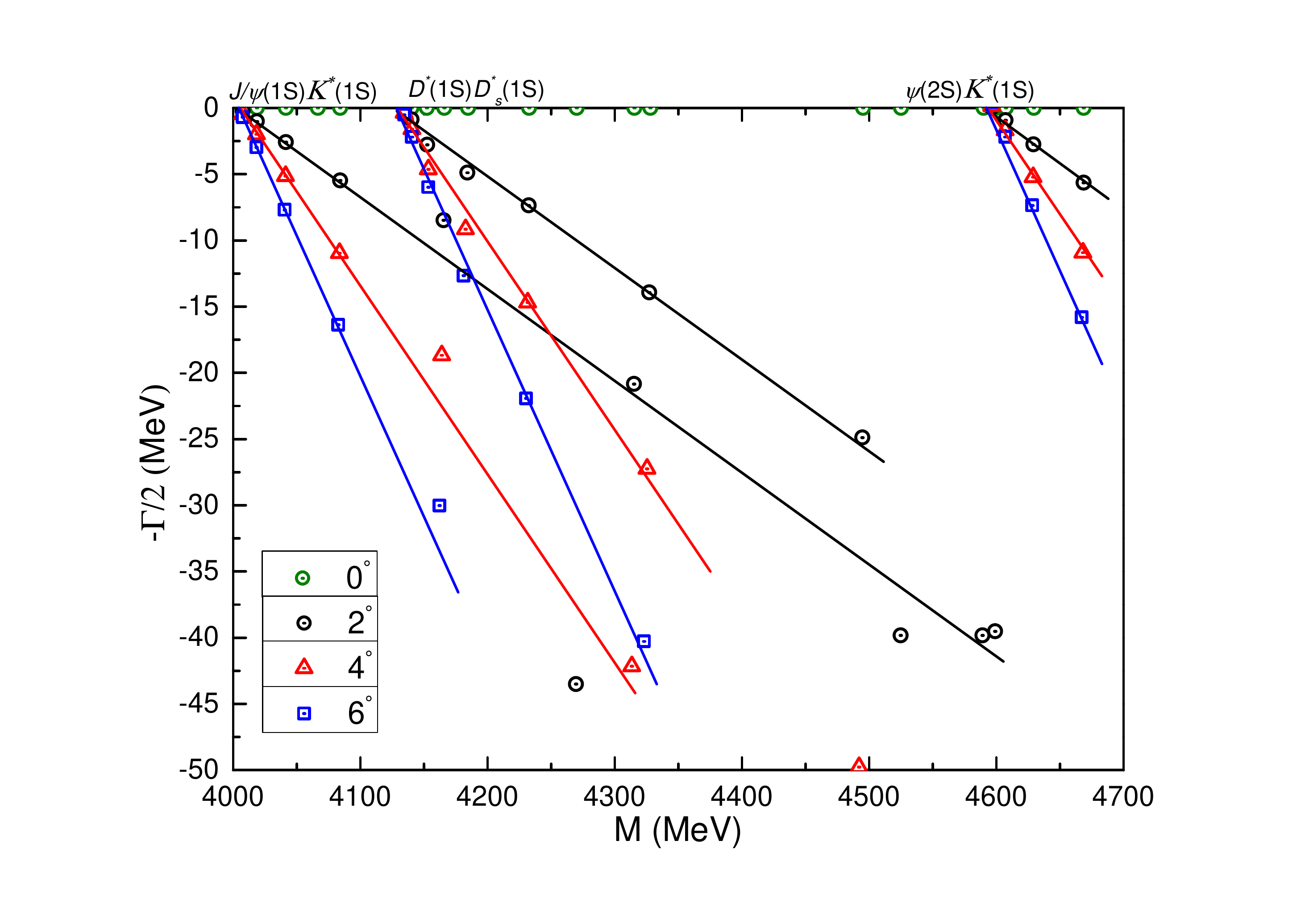}
\caption{\label{PP9} The complete coupled-channels calculation of the $c\bar{c}s\bar{q}$ tetraquark system with $J^P=2^+$ quantum numbers. We use the complex-scaling method of the chiral quark model varying $\theta$ from $0^\circ$ to $6^\circ$.}
\end{figure}

Table~\ref{GDD3} lists the allowed 14 channels in the highest spin state of the $c\bar{c}s\bar{q}$ tetraquark system considered herein. Particularly, there are four meson-meson structures, two diquark-antidiquark structures and 8 K-type ones. In a real-range investigation, Table~\ref{GresultCC3} summarizes the calculated results of these channels. Therein, the physical channels are listed in the first column, and they are indexed in the following one. The theoretical mass of each channel and coupled result in each kind of configuration is listed in the 3rd and 4th column, respectively. Meanwhile, the lowest-lying mass of system in a complete coupled-channels is listed in the last row of Table~\ref{GresultCC3}. Additionally, Figs.~\ref{PP7} to~\ref{PP9} show the coupled-channels results in a complex-range study, and the resonance states are indicated inside circles. When coupling is only considered for the exotic color configurations: the hidden-color channels of dimeson structures, diquark-antidiquark structures and K-type ones, Table~\ref{tab:dis3} presents the exotic resonances whose masses are less than 4.3 GeV, and their inner structures are also analyzed. Further details are discussed in the following.

{\bf Exotic states in $(c\bar{c})(s\bar{q})$ dimeson channels:} The calculated theoretical masses for the $J/\psi K^*$ in both color-singlet and -octet channels are 4004 and 4466 MeV, respectively (see Table~\ref{GresultCC3}). Therefore, the $(J/\psi K^*)^1$ state is of scattering nature and cannot be identified as the $Z_{cs}(4000)$ reported by the LHCb collaboration.

Figure~\ref{PP7} illustrates the distribution of complex energies when a coupled-channels investigation that includes the $J\psi K^*$ dimeson, diquark-antidiquark and K-type structures is performed in the CSM. Within a mass region 4.0$\sim$4.7 GeV, the scattering states of $J/\psi(1S)K^*(1S)$ and $\psi(2S)K^*(1S)$ are well obtained. However, one could also realize that three narrow resonances, circled with green lines, are not sensitive with respect to the variation of the complex angle $\theta$. The calculated masses and widths for these resonances are $(4271, 5.2)$ MeV, $(4553, 4.0)$ MeV and $(4678, 1.6)$ MeV. In analogy with the $\psi(2S)K^*(1S)(4675)$ resonance obtained in the $0^+$ channel, the $\psi(2S)K^*(1S)(4678)$ resonance obtained here could also be related to the announced $X(4630)$, since the spin-parity of the $X(4685)$ state has been experimentally assigned to be $1^+$.

{\bf Exotic states in $(c\bar{q})(s\bar{c})$ dimeson channels:} Considering the $D^* D^*_s$ di-meson channel with quantum numbers $2^+$ shown in Table~\ref{GresultCC3}, the lowest masses of color-singlet and hidden-color structures are 4132 and 4411 MeV, respectively, and thus our conclusion on having scattering states in color-singlet channel remains. Figure~\ref{PP8} shows the distribution of the complex energy dots obtained in a coupled-channels calculation with $D^* D^*_s$ di-meson configurations, diquark-antidiquark structures and K-type arrangements. Particularly, the scattering nature of $D^*(1S)D^*_s(1S)$ is clearly shown in the 4.1$\sim$4.7 energy region. However, two quite close resonance poles are circled in the complex plane of Fig.~\ref{PP8}. Therein, the two resonance states can be identified as $D^*(1S)D^*_s(1S)(4685)$ and $D^*(1S)D^*_s(1S)(4692)$, their widths are 4.8 and 1.6 MeV, respectively. Although the mass of the $D^*(1S)D^*_s(1S)(4685)$ structure extremely coincides with that of the $X(4685)$ state, the spin-parity $2^+$ is different from its experimental assignment. 

\begin{table}[!t]
\caption{\label{tab:dis3} The distance, in fm, between any two quarks of the $J^P=2^+$ $c\bar{c}s\bar{q}$ tetraquark resonance state obtained in all exotic configurations' coupled-channels calculation. These resonances, which masses are below 4.3 GeV, are labeled in the first column.}
\begin{ruledtabular}
\begin{tabular}{ccccccc}
  ~~State & $r_{\bar{c}c}$ & $r_{\bar{c}\bar{q}}$ & $r_{\bar{c}s}$ & $r_{c\bar{q}}$  & $r_{cs}$ & $r_{s\bar{q}}$~~  \\[2ex]
  ~~$Z_{cs}(4145)$& 0.41 & 0.96 & 0.88 & 0.95 & 0.88 & 0.88 ~~ \\
  ~~$Z_{cs}(4276)$& 0.65 & 0.83 & 0.63 & 0.74 & 0.74 & 0.92 ~~ \\
\end{tabular}
\end{ruledtabular}
\end{table}

{\bf The fully coupled-channels case:} As shown in Table~\ref{GresultCC3}, the exotic configurations, which are diquark-antidiquark and K-type arrangements, lie generally in the mass range 4.3$\sim$4.4 GeV. In coupled-channels computations for each kind of configuration, the lowest-lying mass located at 4004 MeV is still the theoretical value of the $(J/\psi K^*)^1$ threshold channel, and this fact is not changed in a fully coupled-channels calculation. Meanwhile, the other coupled-channels structures have higher masses at around 4.3 GeV.

When all of the exotic color structures are considered in a coupled-channels study, Table~\ref{tab:dis3} lists two resonance states whose mass is below 4.3 GeV. The distances between any two quarks are less than 1 fm, and hence they are good candidates of compact tetraquarks with color resonance structures.

In a further step, when the CSM is employed in a fully coupled-channels calculation, Fig.~\ref{PP9} shows the distribution of the scattering states for $J/\psi(1S)K^*(1S)$, $D^*(1S)D^*_s(1S)$ and $\psi(2S)K^*(1S)$ in the mass region 4.0$\sim$4.7 GeV. As in the case of spin-parity $0^+$, no bound state is found and the calculated resonance states in different kinds of coupled-channels studies, \emph{e.g.}, $\psi(2S)K^*(1S)(4678)$, $D^*(1S)D^*_s(1S)(4685)$ and $D^*(1S)D^*_s(1S)(4692)$, etc., are quite unstable decaying easily to the $J/\psi K^*$ and $D^* D^*_s$ meson-meson scattering states.


\section{Summary}
\label{sec:summary}

A systematical investigation of hidden-charm tetraquarks with strange content: $c\bar{c}s\bar{q}$ $(q=u,\,d)$, has been performed within a chiral quark model formalism. The model, which includes the one-gluon exchange, a linear-screened confining and Goldstone-boson exchange interactions between quarks, has been successfully applied to the description of hadron, hadron-hadron and multiquark phenomenology. In particular, the hidden-charm pentaquarks and doubly charmed tetraquark are well predicted in our previous theoretical investigations. Our formulation in real- and complex-scaling method of the theoretical formalism allows us to distinguish three kinds of scattering singularities: bound, resonance and scattering. Furthermore, the meson-meson, diquark-antidiquark and K-type configurations, plus their couplings, are considered for the tetraquark system. Finally, the Rayleigh-Ritz variational method is employed in dealing with the spatial wave functions of the $c\bar{c}s\bar{q}$ tetraquark states, which are expanded by means of the well-known Gaussian expansion method (GEM) of Ref.~\cite{Hiyama:2003cu}.

Our theoretical findings can be summarized as follows.
\begin{itemize}
\item The $Z_{cs}(3985)$ and $Z_{cs}(4000)$ can both be identified as compact $c\bar{c}s\bar{q}$ tetraquark states with $J^P=1^+$, and their sizes are less than 1 fm.
\item The $Z_{cs}(4220)$ is compatible with being a hadronic molecular resonance of either $\eta_c(2S)K(1S)(4255)$ with spin-parity $0^+$ or $\psi(2S)K(1S)(4254)$ and $D^*(1S)D^*_s(1S)(4254)$ with $1^+$ quantum numbers.
\item The $X(4685)$ can be well identified as a hadronic molecular resonance whose structure resembles the $D(1S)D^*_s(2S)(4695)$ arrangement with quantum numbers $J^P=1^+$.
\item An extra exotic state $Z_{cs}(4150)$, which is predicted in other theoretical investigation, can be explained as $D^*(1S)D^*_s(1S)$ resonance in $0^+$ state herein.
\item Several compact $c\bar{c}s\bar{q}$ tetraquark resonances within a mass region 3.8$\sim$4.2 GeV and narrow hadronic molecular resonances, which locate at 4.1$\sim$4.3 GeV and 4.5$\sim$4.6 GeV, are obtained in $0^+$, $1^+$ and $2^+$ states, respectively.
\end{itemize}


\begin{acknowledgments}
Work partially financed by: National Natural Science Foundation of China under Grant Nos. 11535005 and 11775118; the Ministerio Espa\~nol de Ciencia e Innovaci\'on, grant no. PID2019-107844GB-C22; and Junta de Andaluc\'ia under contract no. Operativo FEDER Andaluc\'ia 2014-2020 UHU-1264517, P18-FR-5057 and PAIDI FQM-370.
\end{acknowledgments}


\bibliography{ccqstetraquarks}

\end{document}